\documentclass[10pt,final,journal,letterpaper,twoside,twocolumn]{IEEEtran}
\usepackage{amsmath}
\usepackage{amssymb}
\usepackage{amsfonts}
\usepackage{amscd}
\usepackage{mathrsfs}
\usepackage{graphicx}
\usepackage{color}
\usepackage{url}
\usepackage{bm}
\usepackage{algorithm}
\usepackage{algorithmic}
\usepackage{setspace}

\newtheorem{theorem}{Theorem}
\newtheorem{lemma}{Lemma}
\newtheorem{definition}{Definition}

\newtheorem{remark}{Remark}

\begin{document}

\title{Full Diversity Unitary Precoded Integer-Forcing}
\author{Amin Sakzad and Emanuele Viterbo$^\dagger$
\thanks{$^\dagger$ Amin Sakzad and Emanuele Viterbo are with
Department of Electrical and Computer Systems, Faculty of Engineering,
Monash University, Clayton VIC 3800, Australia.
E-mail: $\tt \{amin.sakzad,emanuele.viterbo$\}@$\tt monash.edu$. This work was performed at the Monash Software Defined Telecommunications (SDT) Lab. Amin Sakzad was supported by the Australian Research Council (ARC) under Discovery grants ARC DP~$130100103$. Emanuele Viterbo was supported by NPRP grant $\mbox{NPRP}5$-$597$-$2$-$241$ from the Qatar National Research Fund (a member of Qatar Foundation). A subset of this work was presented in~\cite{sakzad14-2} at ITW $2014$, Hobart, Tasmania, Australia.}
}

\maketitle
\IEEEpeerreviewmaketitle
\begin{abstract}
We consider a point-to-point flat-fading MIMO channel with channel state information known both at transmitter and receiver. At the transmitter side, a lattice coding scheme is employed at each antenna to map information symbols to independent lattice codewords drawn from the same codebook. Each lattice codeword is then multiplied by a unitary precoding matrix ${\bf P}$ and sent through the channel. At the receiver side, an integer-forcing (IF) linear receiver is employed. We denote this scheme as {\em unitary precoded integer-forcing} (UPIF). We show that UPIF can achieve full-diversity under a constraint based on the shortest vector of a lattice generated by the precoding matrix ${\bf P}$. This constraint and a simpler version of that provide design criteria for two types of full-diversity UPIF. Type I uses a unitary precoder that adapts at each channel realization. Type II uses a unitary precoder, which remains fixed for all channel realizations. We then verify our results by computer simulations in $2\times2$, and $4\times 4$ MIMO using different QAM constellations. We finally show that the proposed Type II UPIF outperform the MIMO precoding X-codes at high data rates.
\end{abstract}
\section{Introduction}
Multiple-input multiple-output (MIMO) channels with multiple antennas appeared in the early $1990$'s as the key technology to achieve high spectral efficiencies in wireless channels. These channels are known to obtain higher capacity gains in comparison to single antenna point-to-point channels. For a MIMO channel matrix ${\bf H}_{n_t\times n_r}$ with $n_t$ transmit antennas and $n_r$ receive antennas, a {\em diversity gain} of at most $n_tn_r$ is achievable. If an encoder/decoder pair gives the maximum possible diversity $n_tn_r$, then the entire system is said to achieve {\em full-diversity}. Let $n_s$ denote the number of symbols transmitted per channel use. If $n_s=\min\{n_r,n_t\}$, then the encoding scheme is said to have {\em full-rate}.

For MIMO channels with channel state information at the receiver (CSIR), carefully designed space-time block-codes (STBC) can provide full-diversity~\cite{TSC}. Algebraic full-diversity full-rate STBCs with large coding gain were later designed in~\cite{BRS,ORBV06}. The design criteria for these codes are known as the {\em rank} and the {\em non-vanishing determinant} criteria~\cite{TSC,ORBV06}. In order to take full advantage from these codes, an optimal maximum likelihood (ML) detector such as a sphere decoder~\cite{ViB}, with high computational complexity should be implemented at the receiver side. Other well-known linear receivers~\cite{Kumar09}, including zero-forcing (ZF) and minimum mean square estimation (MMSE) linear receivers, trade-off error performance for reduced computational complexity. Specific families of full-diversity STBCs were investigated and constructed in~\cite{ZLW1}--\cite{WXYL} for use with linear receivers (ZF and MMSE). In these schemes, full-diversity is obtained at the cost of limited rate.

Precoding schemes that assume CSIR, channel state information at the transmitter (CSIT), and an ML decoder usually make use of a singular value decomposition (SVD)  to convert the MIMO channel into parallel subchannels. Then appropriate precoders employed over these parallel channels to minimize the error probability of the system over different constellation sizes. Some examples are the well-known full-diversity E-$d_{\min}$ precoders for $4$-QAM~\cite{Verigneu08}, and MIMO precoding X- and Y- codes~\cite{saif11}, which are known to provide a diversity order of $d_{n_t,n_r,n_s} = \left(n_r-\frac{n_s}{2}+1\right)\left(n_t-\frac{n_s}{2}+1\right)$,
and the full-diversity precoders introduced and investigated in~\cite{pavan11}. Note that $d_{n_t,n_r,n_s}$ shows a trade-off between full-rate and full-diversity. In one hand, if $n_t=n_r$ and we use a full-rate MIMO precoding X- or Y- code, then the diversity order is
$d_{n_t,n_r,n_s} = \left(\frac{n_t}{2}+1\right)^2 < n_t^2$ (full-diversity).
On the other hand, if we want to ensure full-diversity for MIMO precoding X- or Y- codes, we need $n_s=2$, which means the system has very limited rate. For a detailed comparison between the ML-decoding complexity of the above three schemes, we refer the reader to Table III of~\cite{pavan11}.

On the other hand, in the presence of CIST and having a linear receiver such as MMSE linear receiver at the destination, optimal precoder designs have been  extensively studied (\cite{Sampath01}--~\cite{Palmor03}). The design criteria for these schemes include maximizing the rate or minimizing the error performance. The MIMO linear precoding techniques such as regularized ZF~\cite{Peel05} are alternative approaches that can provide full-diversity in conjunction with linear receivers, see~\cite{Aria14} and references therein. The decoding complexity of such schemes is relatively low in comparison to ML-decoding algorithms but incur some error performance degradation.

In this paper, assuming CSIT and CSIR we propose a full-diversity unitary precoding scheme for $n\times n$ MIMO ($n_r=n_t=n$), where the receiver is equipped with the recently proposed integer-forcing (IF) linear receiver~\cite{zhan12}. In IF framework, the transmitter employs a lattice coding scheme, and sends independent lattice codewords from the same codebook, across different transmit antennas (or {\em layers}). At each receive antenna, we recover an integer linear combination of the lattice codewords depending on the entries of the channel matrix ${\bf H}$. As all the lattice codewords are drawn from the same lattice codebook, the integer linear combinations are themselves guaranteed to be lattice points. To further estimate the original transmitted lattice codewords, a system of linear equations with integer coefficient matrix ${\bf A}$ should be solved at the receiver. In fact, the IF linear receiver exploits the linearity of lattices to first eliminate the noise and then hareness the interference between transmitted data across different layers. The key step underpinning IF is the selection of ${\bf A}$ to approximate the channel matrix ${\bf H}$. Hence, a matrix ${\bf B}$ is needed such that ${\bf A}$ is full rank and ${\bf B}{\bf H} \approx {\bf A}$, with minimum quantization error at high signal-to-noise ratio (${\small \mbox{SNR}}$) values. The problem of finding ${\bf A}$ based on ${\bf H}$ using lattice reduction algorithms is addressed in~\cite{Sakzad14-1, Sakzad13-1}. The integer-forcing linear receiver is also shown to provide a full receive diversity order $n_r=n$ and a full multiplexing gain in~\cite{zhan12, Sakzad14-1}.

Using perfect STBCs like the Golden code~\cite{BRV05,ORBV06}, the authors of~\cite{OrE} have shown that under an IF decoder, a constant gap to capacity is attainable. A perfect STBC code is a linear dispersion space-time code over a Quadrature-Amplitude Modulation (QAM) constellation, which is full-rate, satisfies non-vanishing determinant property, and its generator matrix is a unitary matrix. In this paper, we aim at relaxing the first two constraints of perfect STBCs and keeping the last one only. We analyze the diversity order of a lattice space-time encoding scheme, which is precoded by a unitary matrix and uses an IF linear receiver. First, singular value decomposition (SVD) precoding is used to diagonalize the channel, and then an  optimal precoder matrix is designed. Our optimization criterion is based on maximizing the minimum distance of the input lattice, similar to the approach of~\cite{Bergman08}. We note that differently from~\cite{Bergman08}, our precoder matrices are unitary and our decoder is the IF linear receiver and not the ML receiver. Related works focus on a different design criterion based on maximizing the mutual information (\cite{Lozano06},~\cite{Perez-Cruz10},~\cite{Telatar99}, and~\cite{Payaro09} and references therein).

From the receiver point of view, using IF linear receiver instead of ML decoders in slow-fading channels is advantageous in terms of complexity. For the IF receiver, the lattice reduction algorithm is performed only at the beginning of each quasi-static interval for which the channel ${\bf H}$ is assumed to remain constant over multiple codeword transmissions. Subsequently, for each codeword transmission, only a system of linear equations has to be solved to recover an estimate of that codeword. On the other hand, for the ML decoding, a sphere decoder should be used for each codeword within a quasi-static channel interval.

We summarize the contributions of this paper as follows:
\begin{itemize}
\item{\bf Unitary precoded integer-forcing (UPIF)} -- Considering the availability of CSIT, the SVD decomposition ${\bf W}{\bf \Sigma}{\bf V}^h$ for ${\bf H}_{n\times n}$ is first performed. Using ${\bf V}$ at the transmitter and ${\bf W}$ at the receiver, the channel is reduced to a diagonal channel matrix ${\bf \Sigma}$. An additional unitary precoding matrix ${\bf P}$, is then employed as  ${\bf VP}$. At the destination, the receiver is equipped with an IF linear receiver.
\item{\bf Full-diversity constraint} -- We find an upper bound on the probability of error of UPIF (Theorem~\ref{th:ub}) and show analytically that if ${\bf P}$ satisfies a specific constraint (Theorem~\ref{th:diversity}), the full diversity ($n^2$) can be achieved by our  UPIF. The exact constraint is hard to verify, since it is related to finding the minimum distance of a lattice depending on ${\bf \Sigma}$ and ${\bf P}$. Theorem~\ref{th:diversity1} provides a simpler constraint, based on the non-vanishing {\em minimum product distance} of the lattice generated by ${\bf P}$ and guarantees the full-diversity of UPIF.
\item{\bf Precoder design criteria} -- The constraints on ${\bf P}$ provide the design criteria for full-diversity-achieving precoders for IF MIMO linear receivers. We define two types of full-diversity UPIF schemes: ({\em i}) the precoder matrix ${\bf P}$ depending on the channel matrix singular values (diagonal elements of ${\bf \Sigma}$) is adapted at each quasi-static channel interval ({\em Type I UPIF}), ({\em ii}) a fixed unitary precoder ${\bf P}$ (independent of ${\bf \Sigma}$) is used for all channel realizations ({\em Type II UPIF}). For Type I UPIF,  finding an analytical optimization technique for ${\bf P}$ remains an open problem. For Type II UPIF we propose the use of full-diversity lattice generator matrices with non-vanishing minimum product distance~\cite{OV04}-\cite{site}.
\item{\bf Simulation results} -- Monte-Carlo simulations are presented in support the derived theoretical results. For example, it is shown that   a $2\times 2$ full-diversity algebraic rotation matrix given in~\cite{site}  in a Type II UPIF scheme with a $64$-QAM constellation, can perform as close as $1$dB away from ML decoder for the same code and $2$dB better than MIMO precoding X-codes~\cite{saif11}.
\end{itemize}

{\bf Notation.} The superscripts $(\cdot)^T$ and $(\cdot)^h$ denote transposition and Hermitian transposition, respectively. Let $\mathcal{G}$ and $\mathcal{G}'$ be a group and its subgroup respectively, then $\mathcal{G}/\mathcal{G}'$ denotes the quotient group. The sets $\mathbb{Z}$, $\mathbb{C}$,  $\mathbb{R}$, and $\mathbb{Z}[i]$ denote the ring of rational integers, the field of complex numbers, the field of real numbers, and the ring of Gaussian integers, respectively, where $i^2 = -1$.
Real and imaginary parts of a complex number $z$ are denoted by
$\Re{(z)}$ and $\Im{(z)}$, and $|z|$ and $\mbox{arg}(z)$ are the modulus and the unique phase, respectively. The notation $\|{\bf v} \|$ stands for the Euclidean norm of a vector ${\bf v}\in\mathbb{C}^n$. A $k\times k$ matrix ${\bf X}=[{\bf x}_1^T,\ldots,{\bf x}_k^T]^T$ is formed by stacking the $k-$dimensional row vectors
${\bf x}_1,\ldots, {\bf x}_k$, and ${\bf I}_k$ denotes the $k\times k$ identity matrix. Finally, let ${\bf v}$ be a vector, then its $j$-th entry is represented by $[{\bf v}]_j$.
\section{System Model}\label{Section:SystemModel}
We first review the notion of complex lattices~\cite{Conway83} which are essential for the rest of the paper.
\begin{definition}
Given a set of $d$-dimensional vectors $\left\{{\boldsymbol\ell}_1,\ldots,{\boldsymbol\ell}_k\right\}$ in $\mathbb{C}^d$, a $k$-dimensional {\em complex lattice} $\Lambda$ is the set of points
$$\left\{\sum_{m=1}^kz_m{\boldsymbol\ell}_m\colon z_m\in\mathbb{Z}[i],~{\boldsymbol\ell}_m\in\mathbb{C}^d\right\}.$$
\end{definition}
More briefly we can write $\Lambda=\left\{{\bf z}{\bf L}\colon{\bf z}\in\mathbb{Z}[i]^k\right\}$, where ${\bf L}=[{\boldsymbol\ell}_1^T,\ldots,{\boldsymbol\ell}_k^T]^T$ represents a {\em basis} of the lattice $\Lambda$. In other words, every point ${\bf x}\in \Lambda$ can be represented as a Gaussian integer linear combination of basis vectors. For any point ${\bf x}\in\Lambda$ the
\emph{Voronoi cell} $\mathcal{V}({\bf x})$ is
$$\left\{{\bf v}=\sum_{m=1}^k\alpha_m{\boldsymbol\ell}_m\colon \|{\bf v}-{\bf x}\|\leq\|{\bf v}-{\bf y}\|,~\forall{\bf y}\in\Lambda,~ \alpha_m\in\mathbb{C}\right\}.$$
The Voronoi cell for the origin is denoted by $\mathcal{V}$. Let ${\bf L}$ be a matrix with ${\boldsymbol\ell}_m$ as its rows, then ${\bf L}$ is called the {\em generator matrix} of the lattice $\Lambda_{\bf L}$. Throughout the paper, we only consider full rank lattices where $d=k$. For example $\mathbb{Z}[i]^d$ is a lattice with the identity matrix as generator matrix. Let $\mathcal{S}\subseteq\mathbb{C}^n$, then the linear span of $\mathcal{S}$ is  defined as:
$$\mbox{span}(\mathcal{S}) \triangleq \left\{\sum_{p=1}^qc_p{\bf v}_p\colon q\in\mathbb{N},~{\bf v}_p\in\mathcal{S},~c_p\in\mathbb{C}\right\}.$$
Let $\mathcal{B}_r({\bf 0})=\left\{{\bf x}\in\mathbb{C}^d\colon\|{\bf x}\|\leq r\right\}$,
be the $d$-dimensional ball of radius $r$ centered at ${\bf 0}$ and $\dim(\mathcal{V})$ denotes the dimension of a subspace $\mathcal{V}$ of $\mathbb{C}^d$. For a $d$-dimensional lattice $\Lambda_{\bf L}$, we define the $m$-th successive minima, for $1\leq m \leq d$ as
\begin{equation}
\epsilon_{m}(\Lambda_{\bf L})\triangleq\inf\left\{r\colon\dim\left(\mbox{span}\left(\Lambda_{\bf L}\cap \mathcal{B}_r({\bf 0})\right)\right)\geq m\right\}.
\end{equation}
The $m$-th successive minima of $\Lambda_{\bf L}$ is the infimum radius  $r$ such that there are $m$ independent vectors of $\Lambda_{\bf L}$ in $\mathcal{B}_r({\bf 0})$.

A $d$-dimensional lattice $\Lambda_{\bf L}$ is called {\em full-diversity} if for all distinct ${\bf x},{\bf y}\in\Lambda_{\bf L}$, $[{\bf x}]_m\neq[{\bf y}]_m $ for all $1\leq m \leq d$.
The {\em minimum product distance} of a full-diversity lattice $\Lambda_{\bf L}$ is denoted by $d_{p,\min}(\Lambda_{\bf L})$ (\cite{BVi98}):
\begin{equation}
d_{p,\min}(\Lambda_{\bf L})\triangleq\min_{{\bf x}\in\Lambda_{\bf L}\setminus\{\bf 0\}}\prod_{m} \left|[{\bf x}]_m\right|.
\end{equation}
A {\em lattice code} $\mathcal{C}\subseteq \Lambda$ is a finite set of points of $\Lambda$. One way of constructing lattice codes is through sublattices (nested lattices). A subset $\Lambda'\subseteq\Lambda$ is called a {\em sublattice} if $\Lambda'$ is a lattice itself. Given a sublattice $\Lambda'$, we define the quotient $\Lambda/\Lambda'$ as a lattice code. The fine lattice $\Lambda$ is called the {\em coding lattice} and the coarse lattice $\Lambda'$ is usually referred as the {\em shaping lattice}. The lattice code $\mathcal{C}=\Lambda/\Lambda'$ corresponds to a finite constellation of lattice points carved from the lattice $\Lambda$, within a fixed Voronoi cell of $\Lambda'$. In other words, let $\mathcal{V}'$ be the Voronoi cell of origin for lattice $\Lambda'$, then $\mathcal{C}=\Lambda\cap\mathcal{V}'$. Therefore, the shape of such constellation is governed by $\mathcal{V}'$. For example, the Voronoi region of the $4$-dimensional checkerboard lattice
$$\Lambda'=D_4\triangleq\left\{(x_1,\cdots,x_4)\in\mathbb{Z}^4\colon \sum_{m=1}^4x_i=\mbox{even}\right\},$$
has exactly $24$ faces, with the relevant vectors being the $24$ minimum Euclidian lattice points~\cite{ConwaySloane}. This lattice is a sublattice of $4$-dimensional integer lattice, {\em i.e.} $D_4 \subseteq \mathbb{Z}^4$ and hence so is the scaled lattice
$8D_4 \triangleq \left\{8{\bf x}\colon {\bf x}\in D_4\right\}\subseteq \mathbb{Z}^4$.
The lattice code $\Lambda/\Lambda'=\mathbb{Z}^4/8D_4$ includes all $4$-dimensional integer points inside the Voronoi region of the lattice $8D_4$.

In general, a common choice for the sublattice $\Lambda'$ of $\Lambda$ is the scaled lattice $g\Lambda$ for some integer $g\in\mathbb{Z}$ as in~\cite{Conway83}. As another simpler example, for $\Lambda=\mathbb{Z}[i]^d$ and $\Lambda'=g\mathbb{Z}[i]^d$, we have $\Lambda/\Lambda'=\mathbb{Z}_g[i]$ for which $\mathbb{Z}_g$ is the ring of integers modulo $g$. For $g=2$ the cubic shaped lattice code $\mathbb{Z}_2[i]$ is simply a $4$-QAM constellation. For the rest of this paper we use $\Lambda=\mathbb{Z}[i]^d$ and choose $g$ to be a power of $2$.

We consider a quasi-static flat-fading $n\times n$ MIMO channel, where the channel state information is available at both transmitter and receiver. The channel matrix is denoted by $\bar{\bf H}\in\mathbb{C}^{n\times n}$, where the entries of $\bar{\bf H}$ are i.i.d. complex Gaussian random variables $\sim\mathcal{CN}(0, 1)$. We use an $n$-layer lattice coding scheme, where the information transmitted across different antennas are independent. For $1 \leq m \leq n$, the $m$-th layer is equipped with a lattice encoder $E:\mathcal{R}^{k}\rightarrow\mathbb{C}^{n}$ which maps a message $\bar{\bf s}_m\in \mathcal{R}^k$ over the ring $\mathcal{R}$ into a lattice codeword $\bar{\bf x}_m \in \Lambda/\Lambda' \subset \mathbb{C}^{n}$ in the complex space. The matrix $\bar{\bf X}=[\bar{\bf x}_1^T,\ldots,\bar{\bf x}_n^T]^T$ is actually a space-time codeword, where its rows are all lattice codewords. This lattice space-time codeword will then be precoded using a unitary matrix. The precoding matrix $\bar{\bf U}$ can be derived using the components of the channel matrix $\bar{\bf H}$ and another unitary matrix $\bar{\bf P}$, to be optimized later. Together with $\bar{\bf U}$, the matrix $\bar{\bf X}$ forms a space-time codeword $\bar{\bf U}\bar{\bf X}$ to be sent through the channel.

\begin{figure*}[htb]%
  \begin{center}%
\includegraphics[width=11cm]{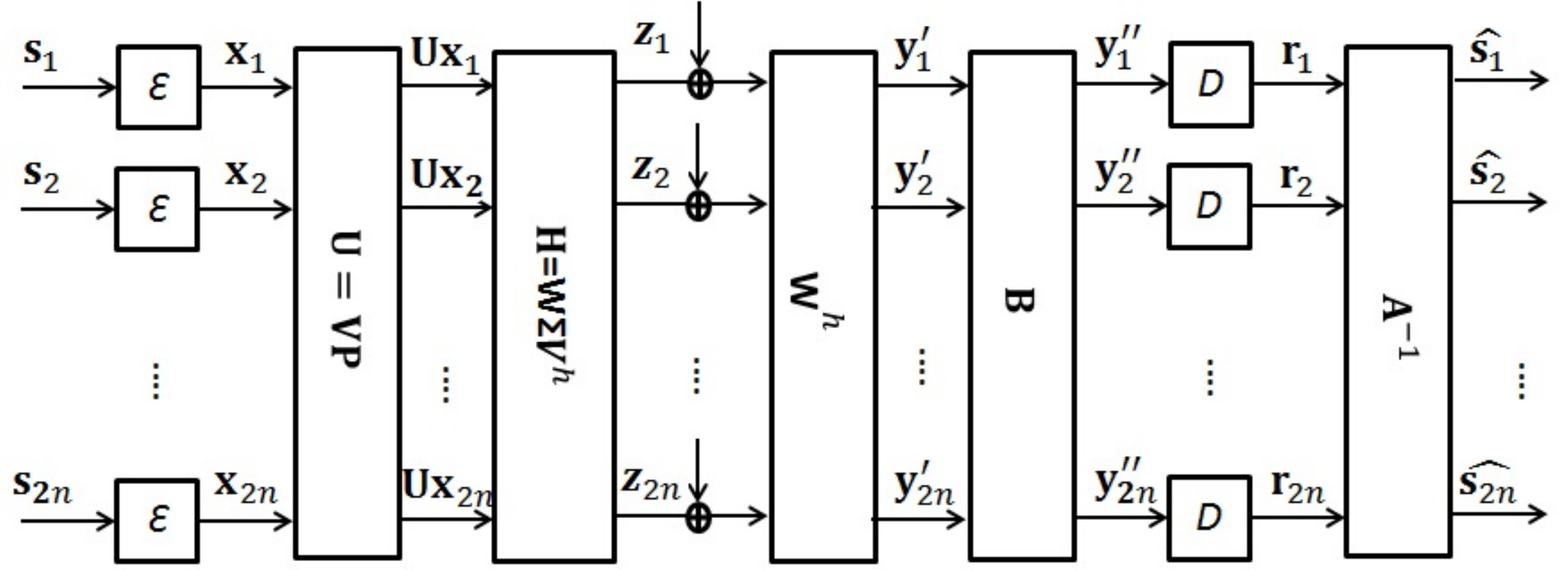}~\caption{\label{fig:bd}
Real Block diagram of unitary precoded integer-forcing.}
  \end{center}
\end{figure*}

Let $\bar{\bf H}=\bar{\bf W}\bar{\bf \Sigma}{\bar{\bf V}}^h$ be the singular value decomposition (SVD) of the channel matrix where $\bar{\bf W},\bar{\bf V}\in\mathbb{C}^{n\times n}$ are two unitary matrices and $\bar{\bf \Sigma}$ is a diagonal matrix given by $\bar{\bf \Sigma}=\mbox{diag}(\bar{\sigma_1},\ldots,\bar{\sigma_n})$ with $\bar{\sigma_1}\geq\cdots\geq\bar{\sigma_n}$ all in $\mathbb{R}$. A unitary precoder matrix
\begin{equation}~\label{eq:precoder}
\bar{\bf U} = \bar{\bf V}\bar{\bf P}
\end{equation}
is then employed at the transmitter where $\bar{\bf P}\in\mathbb{C}^{n\times n}$ is a unitary matrix that needs to be optimized. If $\bar{\bf X}=[\bar{\bf x}_1^T,\ldots,\bar{\bf x}_n^T]^T$ denotes the matrix of transmitted vectors, the received signal $\bar{\bf Y}$ is given by
\begin{equation}~\label{eq:model1}
\bar{\bf Y} = \sqrt{\rho}\cdot\bar{\bf H}\bar{\bf U}\bar{\bf X}+\bar{\bf Z},
\end{equation}
where $\rho = \frac{\mbox{\small SNR}}{n}$ denotes the average signal-to-noise ratio at each receive antenna and the entries of $\bar{\bf Z}$ are i.i.d. distributed as $\mathcal{CN}(0, 1)$.
Let ${\bar{\bf M}}$ be the generator matrix of a complex lattice $\Lambda_{{\bar{\bf M}}}$, then by using the standard conversion from complex lattice to the equivalent real one, we have the real lattice generator matrix as:
$${\bf M} = \left(\begin{array}{cc}
\Re(\bar{\bf M})&\Im(\bar{\bf M})\\
-\Im(\bar{\bf M})&\Re(\bar{\bf M})\\
\end{array}\right).$$
With the above transformation, \eqref{eq:model1} can be written as
\begin{equation}~\label{eq:model3}
{\bf Y} = \sqrt{\rho}\cdot{\bf H}{\bf U}{\bf X}+{\bf Z},
\end{equation}
where all the matrices ${\bf Y}, {\bf H}, {\bf U}, {\bf X}$, and ${\bf Z}$ are in $\mathbb{R}^{2n\times 2n}$. A suitable block diagram is then as Fig.~\ref{fig:bd}. Upon receiving ${\bf Y}$ at the destination, we multiply it by ${\bf W}^h$ to get ${\bf Y}'\triangleq{\bf W}^h{\bf Y}$.
Substituting ${\bf U}$ from \eqref{eq:precoder} into \eqref{eq:model3} the channel can be modeled as:
\begin{equation}~\label{eq:eqmodel}
{\bf Y}' = \sqrt{\rho}\cdot{\bf \Sigma}{\bf P}{\bf X}+{\bf Z}',
\end{equation}
where ${\bf Z}'={\bf W}^h{\bf Z}$. Note that ${\bf Z}'$ continues to be an i.i.d. complex Gaussian matrix with entries $\sim\mathcal{CN}(0, 1)$ because ${\bf W}$ is unitary.

At this point, the receiver employs integer-forcing. That is a linear receiver architecture that creates an effective integer-valued channel matrix. The entries of this effective channel matrix are used to recover integer combinations of the codewords. This is in contrast to the other linear receivers and ML-based decoders, which attempt to recover the transmitted codewords directly. A feature of integer-forcing linear receiver is in using the same lattice codes at both the transmitter as well as the receiver. In this framework, the sender uses a layered transmission scheme, and transmits independent lattice codewords simultaneously across the layers. At the receiver side, each layer is allowed to decode an integer linear combination of transmitted lattice codewords. Since any integer linear combination of lattice points is another lattice point, the decoded point will be another lattice point. Assuming that the effective channel matrix is full rank, these integer combinations can then be solved for the original lattice codewords.

The goal of integer-forcing linear receiver is to project ${\bf \Sigma}{\bf P}$ (by left multiplying it with a receiver filtering matrix ${\bf B}$) onto a non-singular integer matrix ${\bf A}$. In order to uniquely recover the information symbols, the matrix ${\bf A}$ must be invertible over the ring $\mathcal{R}$. Thus, we have
\begin{equation}~\label{eq:LRmodel}
{\bf Y}''={\bf B}{\bf Y}'=\sqrt{\rho}\cdot{\bf B}{\bf \Sigma}{\bf P}{\bf X}+{\bf B}{\bf Z}'.
\end{equation}
For the IF receiver~\cite{zhan12} formulation, the model is given by
\begin{eqnarray}
{\bf Y}''&=&\sqrt{\rho}\cdot{\bf A}{\bf X}+\underbrace{\sqrt{\rho}\cdot({\bf B}{\bf \Sigma}{\bf P}-{\bf A}){\bf X}}_{\mbox{quantization noise term}}+{\bf B}{\bf Z}'\nonumber\\
&\triangleq&\sqrt{\rho}\cdot{\bf A}{\bf X}+{\bf E}~\label{eq:IFmodel},
\end{eqnarray}
where $\sqrt{\rho}\cdot{\bf A}{\bf X}$ is the desired signal component, and the effective noise is
\begin{equation}~\label{eq:effectivenoise}
\sqrt{\rho}\cdot({\bf B}{\bf \Sigma}{\bf P}-{\bf A}){\bf X}+{\bf B}{\bf Z}'.
\end{equation}
We further denote the effective noise term along the $m$-th layer by ${\bf e}_m$ and by stacking these row vectors, we get ${\bf E}$.
The average energy of the effective noise along the $m$-th row of ${\bf Y}''$ is defined as
\begin{equation}~\label{quntizederrplusnoise}
G({\bf a}_m,{\bf b}_m)\triangleq \rho\|{\bf b}_m{\bf \Sigma}{\bf P}-{\bf a}_m\|^2 + \|{\bf b}_m\|^2,
\end{equation}
where ${\bf a}_m$ and ${\bf b}_m$ denote the $m$-th row of ${\bf A}$ and $\bf{ B}$, respectively. Note that in order to increase the $m$-th layer effective signal-to-noise ratio~\cite{zhan12}
\begin{equation}~\label{eq:effSNR}
{\small \mbox{SNR}}_{\tiny \mbox{eff}} = \frac{\rho}{G({\bf a}_m,{\bf b}_m)}
\end{equation}
the term $G({\bf a}_m,{\bf b}_m)$ has to be minimized, for each $m$, by appropriately selecting the matrices ${\bf A}$ and ${\bf B}$. We refer to the above signal model as {\em unitary precoded integer-forcing}. Note that the signal strength in the numerator of \eqref{eq:effSNR} is simply $\rho$ (instead of $\rho\|{\bf a}_m\|^2$) due to the fact that ${\bf a}_m^T{\bf x}$ is itself a codeword, which is decodable up to an effective noise variance of $\rho$. Specifically, for lattice codes, it is often easier to think of each codeword as surrounded by a ball of radius $\sqrt{n\rho}$. If the noise falls within this ball, then decoding is successful. The received signal-strength viewpoint is more appropriate for i.i.d. random coding methods, where a joint typicality analysis can be applied.
\subsection{Decoding Complexity Advantage of UPIF}
We highlight the decoding complexity advantage of UPIF in comparison to ML decoder techniques such as sphere decoder~\cite{ViB} in slow-fading MIMO channels. In these channels, the phases and the amplitudes of the channel coefficients can be assumed approximately constant over a long period. One such period is called a {\em quasi-static channel interval} or simply an interval. Let the channel coefficients remain unchanged for a time period $t$, which is much larger than the codeword length $2n$, so that many codewords can be transmitted in this interval. For example, let $m=\frac{t}{2n}\in\mathbb{Z}$ be the number of feasible codeword transmissions within one such quasi-static channel interval. In IF, the matrices ${\bf A}$ and ${\bf B}$ are computed once, based on ${\bf H}$ and estimated $\rho$ in the first codeword transmission. In the subsequent $m-1$ codeword transmissions within the same interval, only a system of linear equations, with integer coefficients ${\bf A}$, will be solved to recover the original information symbols at the receiver. Referring to Table II of~\cite{Sakzad14-1}, the overall complexity of using IF in an interval is dominated by the complexity of finding the matrix ${\bf A}$ as the $m$ systems of $2n$ linear equations with $2n$ unknown variables can be solved using Gauss-Jordan elimination process with computational complexity of order $m\times(2n)^3$. This is in contrast to ML decoding algorithms in slow-fading MIMO channels. In these scenarios, an algorithm such as sphere decoder should be employed for each and all $m$ codeword transmissions inside one quasi-static channel interval. Therefore, the overall computational complexity for an interval is $m$ times the complexity of the sphere-decoder algorithm~\cite{ViB}. This makes IF linear receiver computationally more efficient than ML decoding algorithms in slow-fading channels.
\section{Diversity Analysis}\label{Section:DiversityAnalysis}
We first recall the definition of diversity and a weak version of Woodbury identity~\cite{Woodbury}, which are used in the rest of this Section.
\begin{definition}
In an $2n\times 2n$ MIMO system and at a high signal-to-noise ration $\rho$, if the average probability $P_e$, that a transmitted symbol vector is wrongly decoded, is approximated by
$\left(c.\rho\right)^{-\delta}$, then $\delta$ is called the {\em diversity order} (or {\em diversity gain}) and $c$ is called the {\em coding gain}. For a MIMO system with precoding, if $\delta=(2n)^2$, then, we
say that the precoder achieves full-diversity order.
\end{definition}
\begin{lemma}[Woodbury Identity]~\label{lem:WoodburyID}
Let ${\bf M}_1$ and ${\bf M}_2$ be two $d\times d$ invertible matrices, then the following identity holds:
$${\bf I}_d -{\bf M}_1 \left({\bf I}_{d}+{\bf M}_2{\bf M}_1 \right)^{-1} {\bf M}_2 = \left({\bf I}_d+{\bf M}_1{\bf M}_2 \right)^{-1}.$$
\end{lemma}
We now proceed to analyze the diversity order of UPIF. The effective noise in \eqref{eq:effectivenoise} is not Gaussian distributed due to the quantization noise term. Note that matrix $\left({\bf I}_{2n}+\rho\cdot{\bf \Sigma}^h{\bf \Sigma}\right)^{-1}$ is a positive definite matrix as it is a diagonal matrix with positive entries, hence it admits a Cholesky decomposition
\begin{equation}~\label{eq:chol}
\left({\bf I}_{2n}+\rho\cdot{\bf \Sigma}^h{\bf \Sigma}\right)^{-1}={\bf L}{\bf L}^h.
\end{equation}
The optimum value of ${\bf b}_{m}$ that minimizes \eqref{quntizederrplusnoise} given ${\bf a}_{m}$ is~\cite{zhan12}
\begin{equation}~\label{eq:bm}
{\bf b}_{m} = \rho\cdot {\bf a}_{m}{\bf \Sigma}{\bf P}^{h}\left({\bf I}_{2n} + \rho\cdot{\bf \Sigma}{\bf P}\left({\bf \Sigma}{\bf P}\right)^{h}\right)^{-1}.
\end{equation}
By defining
$${\bf S}\triangleq{\bf I}_{2n} + \rho\cdot{\bf \Sigma}{\bf P}\left({\bf \Sigma}{\bf P}\right)^{h},$$
we re-write \eqref{eq:bm} as
\begin{equation}~\label{eq:optimalb}
{\bf b}_{m} = \rho\cdot {\bf a}_{m}{\bf \Sigma}{\bf P}^{h}{\bf S}^{-1}.
\end{equation}
Let
\begin{equation}~\label{eq:defineLp}
{\bf L}_p\triangleq{\bf P}^h{\bf L},
\end{equation}
then the average energy of the effective noise term along the $m$-th layer is given by
\begin{eqnarray}
G({\bf a}_m,{\bf b}_m)&=&\!\!\rho\|{\bf b}_m {\bf \Sigma}{\bf P}-{\bf a}_m\|^2 +\|{\bf b}_m\|^2\nonumber\\
&=&\!\!\rho\cdot{\bf a}_{m}({\bf I}_{2n}-\left({\bf \Sigma}{\bf P}\right)^h{\bf S}^{-1}{\bf \Sigma}{\bf P}){\bf a}_{m}^h\label{eq:new1}\\
&=&\!\!\rho\cdot{\bf a}_{m}\left({\bf I}_{2n}+\rho\cdot\left({\bf \Sigma}{\bf P}\right)^h{\bf \Sigma}{\bf P}\right)^{-1}{\bf a}_{m}^h\label{eq:new2}\\
&=&\!\!\rho\cdot{\bf a}_{m}{\bf P}^h\left({\bf I}_{2n}+\rho\cdot{\bf \Sigma}^h{\bf \Sigma}\right)^{-1}{\bf P}{\bf a}_{m}^h\nonumber\\
&=&\!\!\rho\cdot{\bf a}_{m}{\bf P}^h{\bf L}{\bf L}^h{\bf P}{\bf a}_{m}^h\label{eq:new3}\\
&=&\!\!\rho\cdot{\bf a}_{m}{\bf L}_p{\bf L}_p^h{\bf a}_{m}^h,\label{eq:new4}
\end{eqnarray}
where \eqref{eq:new1} holds because of \eqref{eq:optimalb} and equation (9) in~\cite{Sakzad14-1}, \eqref{eq:new2} is true because of the Woodbury identity in Lemma~\ref{lem:WoodburyID} for
$$\begin{array}{lll}
d=2n, & {\bf M}_1 = \sqrt{\rho}\cdot\left({\bf \Sigma}{\bf P}\right)^h, & {\bf M}_2 = \sqrt{\rho}\cdot{\bf \Sigma}{\bf P},
\end{array}$$
and \eqref{eq:new3} and \eqref{eq:new4} are obtained based on \eqref{eq:chol} and \eqref{eq:defineLp}, respectively.
We denote the probability of error for decoding the $m$-th layer in the integer lattice $\mathbb{Z}^{2n}$ by $P_{e}\left(m, {\bf \Sigma}{\bf P}, \mathbb{Z}^{2n}\right)$. It follows that
\begin{equation}~\label{eq:mthlayerPe}
P_{e}\left(m, {\bf \Sigma}{\bf P}, \mathbb{Z}^{2n}\right) = \mbox{Pr}\left(\left|{\bf e}_m\right| \geq \frac{\sqrt{\rho}}{2}\right),
\end{equation}
where ${\bf e}_m$ is the $m$-th row of ${\bf E}$ defined in \eqref{eq:IFmodel}.

The following Theorem provides an upper bound on $P_{e}\left(m, {\bf \Sigma}{\bf P}, \mathbb{Z}^{2n}\right)$.
\begin{theorem}[Upper Bound on Probability of Error]~\label{th:ub}
For all $1\leq m \leq 2n$, the term $P_{e}\left(m, {\bf \Sigma}{\bf P}, \mathbb{Z}^{2n}\right)$ as in \eqref{eq:mthlayerPe}, is upper bounded as
\begin{equation}\label{P_e_bound_2}
P_{e}\left(m, {\bf \Sigma}{\bf P}, \mathbb{Z}^{2n}\right) \leq \exp\left(-c\epsilon_{2n-m+1}^{2}(\Lambda_{{\bf L}_p^{-1}})\right),
\end{equation}
where $c$ is a constant independent of $\rho$ and $\epsilon_{2n-m+1}^{2}(\Lambda_{{\bf L}_p^{-1}})$ is the $(2n-m+1)$-th successive minima of the lattice with generator matrix ${\bf L}_p^{-1}$ defined in \eqref{eq:new4}.
\end{theorem}
\begin{IEEEproof}
The proof is given in Appendix~\ref{app:upperboundproof}.
Note that in the proof we assume the use of dithering which makes ${\bf x}_m$ a random variable distributed uniformly within the shaping region.
\end{IEEEproof}
We are interested in $P_{e}\left(2n, {\bf \Sigma}{\bf P}, \mathbb{Z}^{2n}\right)$ which is an upper bound for $P_{e}\left(m, {\bf \Sigma}{\bf P}, \mathbb{Z}^{2n}\right)$, for all $1\leq m \leq 2n$. We define the error probability for unitary precoded integer forcing over $\mathbb{Z}^{2n}$ as
\begin{equation}~\label{def:upper1}
P_{e}\left({\bf \Sigma}{\bf P}, \mathbb{Z}^{2n}\right) \triangleq P_{e}\left(2n, {\bf \Sigma}{\bf P}, \mathbb{Z}^{2n}\right).
\end{equation}
Based on \eqref{P_e_bound_2}, we have
\begin{equation}~\label{def:upper2}
P_{e}({\bf \Sigma}{\bf P}, \mathbb{Z}) \leq \exp\left(-c\epsilon_{1}^{2}(\Lambda_{{\bf L}_p^{-1}})\right).
\end{equation}
Let the average probability
$$P_{e} = \mathbb{E}_{{\bf H}}\left(P_{e}({\bf \Sigma}{\bf P}, \mathbb{Z})\right),$$
where the expectation is taken over all channel matrices ${\bf H}$.
\begin{theorem}~\label{th:diversity}
Let the precoding matrix ${\bf P}$ be such that $[{\bf P}{\bf v}]_1\neq 0$, where ${\bf v}\in\mathbb{Z}^{2n}$ is the vector satisfying $\epsilon_1^2(\Lambda_{{\bf L}_p^{-1}}) = \|{\bf L}_p^{-1}{\bf v}\|^2$, then the unitary precoded integer-forcing explained through \eqref{eq:model3} to \eqref{quntizederrplusnoise} achieves full-diversity $(2n)^2$.
\end{theorem}
\begin{IEEEproof}
The proof is given in Appendix~\ref{app:diversityproof}.
\end{IEEEproof}
Based on the proof of the above theorem, to achieve full-diversity, all we need is to select a unitary precoder matrix ${\bf P}$ such that if ${\bf v}\in \mathbb{Z}^{2n}$ be the vector, which arises to the minimum distance of the lattice $\Lambda_{{\bf L}_p^{-1}}$, then it satisfies $\left[{\bf P}{\bf v}\right]_1\neq0$. However, enforcing this condition seems to be unpractical because it involves computation of $d_{\min}\left(\Lambda_{{\bf L}_p^{-1}}\right)$ which is an NP-hard problem. For that reason, we choose to work with a stronger condition that is $d_{p,\min}(\Lambda_{\bf P})\neq0$. It is clear that if $d_{p,\min}(\Lambda_{\bf P})\neq0$ then $\left[{\bf P}{\bf v}\right]_1\neq0$ for every vector ${\bf 0}\neq{\bf v}\in\mathbb{Z}^{2n}$.

\begin{theorem}~\label{th:diversity1}
Let the precoding matrix ${\bf P}$ be such that $d_{p,\min}(\Lambda_{\bf P})\neq0$, then the achievable diversity of the unitary precoded integer-forcing explained through \eqref{eq:model3} to \eqref{quntizederrplusnoise} is $(2n)^2$.
\end{theorem}
According to the above theorem, if lattice codes along with a non-zero minimum product distance precoder are used at the transmitter of a MIMO channel and integer-forcing is employed at the receiver, full (transmit and receive) diversity can be achieved. We next provide two different design criteria for obtaining unitary precoders suitable for integer-forcing linear receivers.
\section{Optimal Design of Full-Diversity Unitary Precoders}\label{Section:OptimalDesignofFull-DiversityIFPrecodersandCodes}
Based on the above two theorems, we next design classes of full-diversity unitary precoders for integer-forcing linear receivers. We first propose a method of designing optimal precoders based on Theorem~\ref{th:diversity} and then we provide our optimal precoders based on Theorem~\ref{th:diversity1}.
\subsection{Design of Type I UPIF}
Based on \eqref{def:upper2} and Theorem~\ref{th:diversity}, the optimal Type I UPIF is as follows:
\begin{equation}~\label{PoptPrecoder}
{\bf P}_{\tiny \mbox{1,opt}} = \arg\max_{{\bf P}\in\mathcal{O}_{2n}}\min_{\substack{{\bf v}\in\mathbb{Z}^{2n}\setminus\{\bf 0\}\\ [{\bf P}{\bf v}]_1\neq0}} \|{\bf L}^{-1}{\bf P}{\bf v}\|^2,
\end{equation}
where $\mathcal{O}_{2n}$ is the orthogonal group of all $2n\times 2n$ orthogonal matrices with matrix multiplication operation. In other words, we should design a precoder matrix ${\bf P}$ such that the minimum distance of the lattice $\Lambda_{{\bf L}_p^{-1}}$ is maximized, while $[{\bf P}{\bf v}]_1\neq0$ for ${\bf v}\in\mathbb{Z}^{2n}$ and $\epsilon_1^2(\Lambda_{{\bf L}_p^{-1}}) = \|{\bf L}_p^{-1}{\bf v}\|^2$.
\begin{remark}
If we relax the constraint on ${\bf P}\in\mathcal{O}_{2n}$ to be $\mbox{Tr}\left({\bf P}{\bf P}^T\right)\leq\rho_0$, where $\mbox{Tr}(.)$ denotes the trace of a matrix, and remove the constriant $[{\bf P}{\bf v}]_1\neq0$, we get a more general optimization problem, which is the main subject of study in~\cite{Kapetanovic13}. The optimal solution for this problem is derived to be the hexagonal lattice $\mathcal{A}_2$~\cite{conway} for $2\times 2$ real MIMO channels and Schlafli lattice $\mathcal{D}_4$~\cite{conway} for $2\times 2$ complex MIMO channels. However, the generator matrix of these lattices are not unitary. In this paper, we focus on solving the optimization problem in \eqref{PoptPrecoder}, which is slightly different from that of~\cite{Kapetanovic13}.
\end{remark}

The problem stated in \eqref{PoptPrecoder} seems to be a hard combinatorial problem, which we could not solve analytically. However, a systematic approach of finding ${\bf P}_{\tiny \mbox{1,opt}}$ is to perform an exhaustive search in $\mathcal{O}_{2n}$. A convenient parametrization of $\mathcal{O}_{2n}$ using $(2n)^2$ parameters is presented in~\cite{Polish}. Such a parametrization uses the fact that an orthogonal transformation ${\bf P}$ can be composed from elementary orthogonal transformations in two-dimensional sub-spaces. The induced matrices using the algorithm given in~\cite{Polish} will be uniformly distributed with respect to Haar measure.

\begin{figure}[t]%
\begin{center}
\includegraphics[width=9cm]{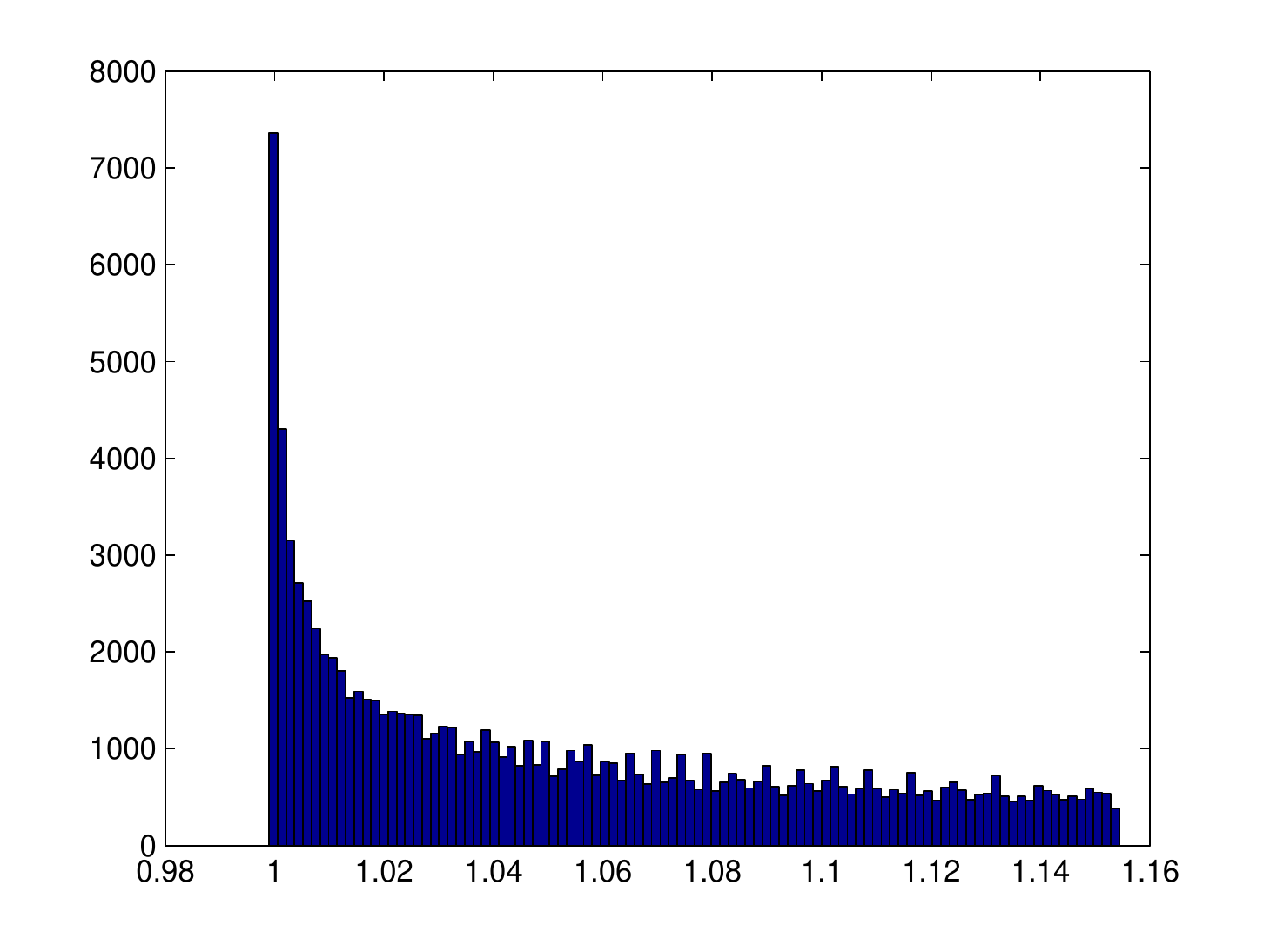}~\caption{\label{fig:histcodinggain}The histogram of $\gamma(\Lambda_{{\bf L}^{-1}_p})$ in a $2\times 2$ complex MIMO Channel using real Type I UPIF.}
\end{center}
\end{figure}

For the $2\times 2$ MIMO system, since we imposed the orthogonality constraint on ${\bf P}$, we parameterize it using a single angle $\theta$ as follows:
\begin{equation}~\label{ParametrizedPrecoder}
{\bf P}(\theta) = \left(\begin{array}{cc}
\cos\theta &\sin\theta\\
-\sin\theta & \cos\theta\end{array}
\right),
\end{equation}
\begin{figure}[t]%
\begin{center}
\includegraphics[width=9cm]{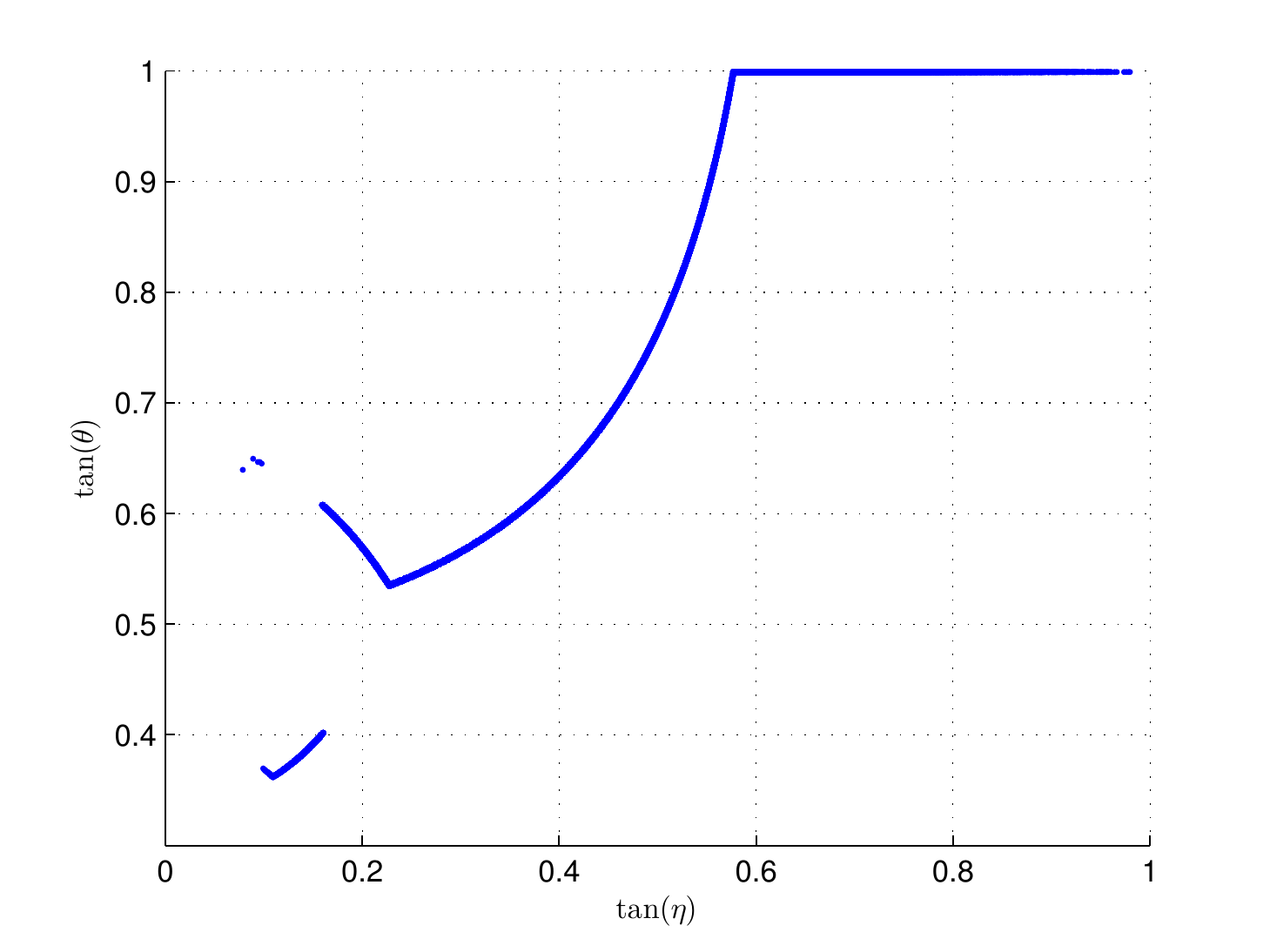}~\caption{\label{fig:tanetavstantheta}The variation of $\tan\theta$ based on the variation of $\tan\eta$ in a $2\times 2$ complex MIMO Channel using real Type I UPIF.}
\end{center}
\end{figure}
It can be easily seen by the symmetry of integers that it suffices to consider $\theta\in[0,\pi/4]$ for the maximization in \eqref{PoptPrecoder}. Hence, we numerically search for
\begin{equation}~\label{PoptPrecoderReal}
{\bf P}^{(\mathbb{R})}_{\tiny \mbox{1,opt}} = \arg\max_{{\bf P}(\theta)\in\mathcal{O}_{2}}\min_{[{\bf P}(\theta){\bf v}]_1\neq0} \|{\bf L}^{-1}{\bf P}(\theta){\bf v}\|^2,
\end{equation}
where $\mathcal{O}_{2}$ is the group of all $2\times 2$ orthogonal matrices with matrix multiplication operation. This exhaustive search can be done by discretizing $\theta\in[0,~\pi/4]$ using fine steps of $0.001$ radians and consequently finding the minimum distance of $\Lambda_{{\bf L}_p^{-1}}$ using Gauss reduction algorithm~\cite{Vallee07}. The $2\times 2$ real precoding matrix, which results in the highest minimum distance of the corresponding lattice $\Lambda_{{\bf L}_p^{-1}}$ is considered as a $2\times 2$ real approximately optimal percoder matrix ${\bf P}^{(\mathbb{R})}_{\tiny \mbox{1,opt}}$ to be employed for both real and complex $2\times 2$ UPIF. Using \eqref{eq:new3}, the matrix ${\bf L}^{-1}$ is given by
$${\bf L}^{-1} = \left(\begin{array}{cc}
\sqrt{1+\rho\sigma_1^2}&0\\
0 & \sqrt{1+\rho\sigma_2^2}\end{array}
\right)=\xi\left(\begin{array}{cc}
\cos\eta&0\\
0 & \sin\eta\end{array}
\right),$$
where $\xi = \sqrt{\left(\sqrt{1+\rho\sigma_1^2}\right)^2 + \left(\sqrt{1+\rho\sigma_2^2}\right)^2}=\sqrt{2 + \rho(\sigma_1^2+\sigma_1^2)}$
and
$$\eta=\tan^{-1}\left(\frac{\sqrt{1+\rho\sigma_2^2}}{\sqrt{1+\rho\sigma_1^2}}\right).$$
It follows that ${\bf L}_p^{-1}$ is equal to
$${\bf L}^{-1}{\bf P}(\theta)=\xi\left(\begin{array}{cc}
\cos\eta&0\\
0 & \sin\eta\end{array}
\right)\left(\begin{array}{cc}
\cos\theta &\sin\theta\\
-\sin\theta & \cos\theta\end{array}
\right).$$
In Fig.~\ref{fig:tanetavstantheta}, we plot the variation of $\tan\theta$ based on $\tan\eta$ for $10^5$ samples of $2\times 2$ complex MIMO systems employing real $2\times 2$ precoders ${\bf P}(\theta)$. In particular,
$$\tan\eta=\frac{\sqrt{1+\rho\sigma_2^2}}{\sqrt{1+\rho\sigma_1^2}}=\frac{1}{\beta_{{\bf L}_p^{-1}}},$$
where $\beta_{{\bf L}^{-1}_p}$ is the condition number of ${\bf L}^{-1}_p$. For large values of $\rho$, $\tan\eta\approx\frac{\sigma_2}{\sigma_1}=\frac{1}{\beta_{\bf H}}$, where $\beta_{\bf H}$ is the condition number of the channel ${\bf H}$. The generator matrix of the lattice $\Lambda_{{\bf L}^{-1}_p}$ can be considered now as the equivalent channel matrix. It can be seen that if the condition number is less than $\sqrt{3}$, then the best precoder obtains by putting $\theta=\pi/4$.
\begin{figure}[t]%
\begin{center}
\includegraphics[width=9cm]{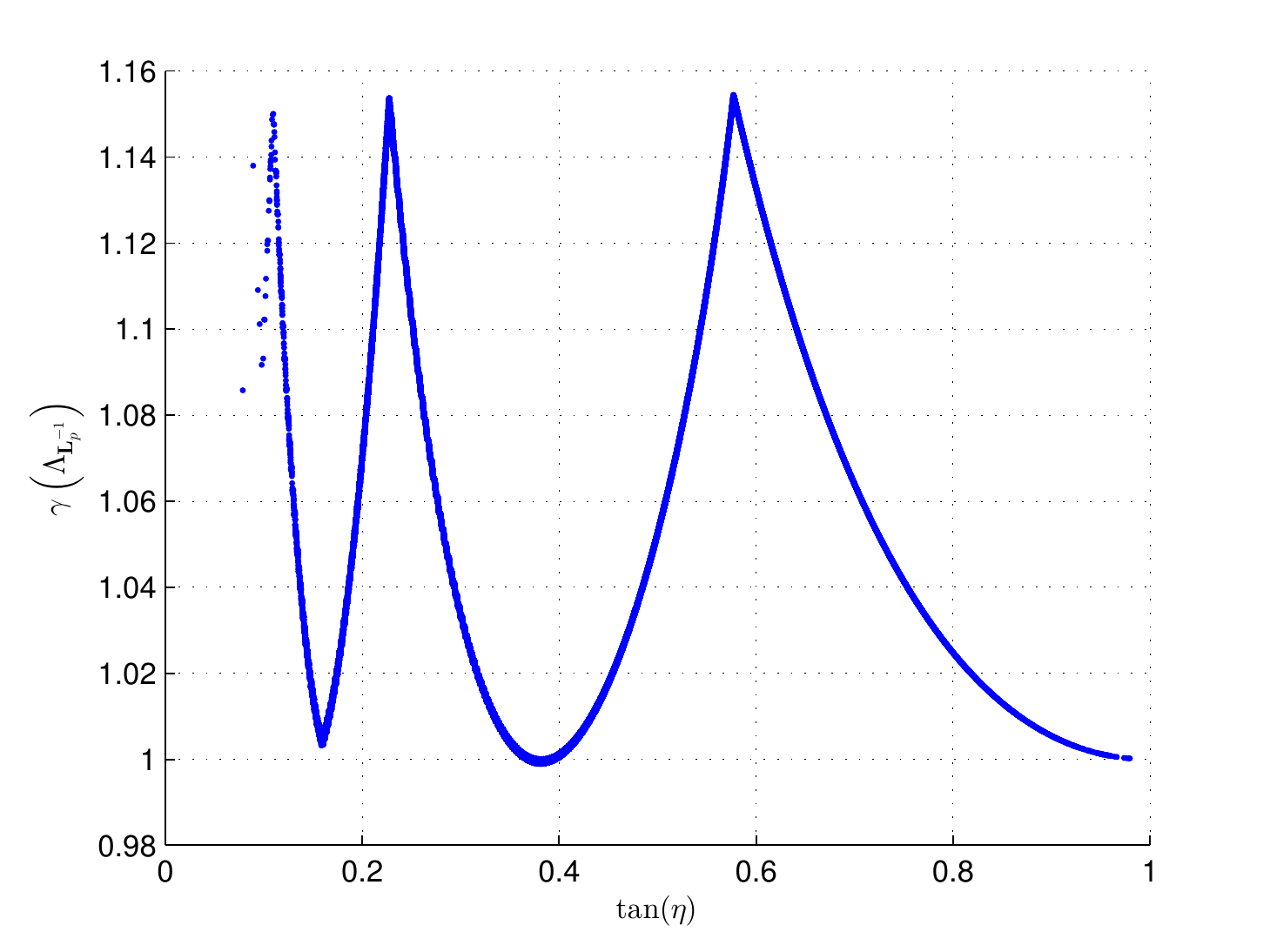}~\caption{\label{fig:tanetavscodinggain}The variation of $\gamma(\Lambda_{{\bf L}^{-1}_p})$ based on the variation of $\tan\eta$ in a $2\times 2$ complex MIMO Channel using real Type I UPIF.}
\end{center}
\end{figure}
In Fig.~\ref{fig:tanetavscodinggain}, we plot the variation of the coding gain of the lattice $\Lambda_{{\bf L}^{-1}_p}$ versus $\tan\eta$. The coding gain formula is:
\begin{equation}~\label{CodingGainParametrizedPrecoder}
\gamma(\Lambda_{{\bf L}^{-1}_p}) = \frac{\epsilon_1^2(\Lambda_{{\bf L}^{-1}_p})}{\det\left({\bf L}^{-1}_p\right)^{\frac{2}{2n}}}.
\end{equation}
The coding gain measures the increase in density of $\Lambda_{{\bf L}^{-1}_p}$ over the integer lattice $\mathbb{Z}^{2n}$ with $\gamma\left(\mathbb{Z}^{2n}\right)=1$. Based on Fig.~\ref{fig:tanetavscodinggain}, we can safely say that the equivalent precoded channel matrix ${\bf L}^{-1}_p$ is at least perform as good as an integer $2\times 2$ channel matrix. In other words, the precoder matrix will change the bad channels to equivalent channels as good as an integer one. Fig.~\ref{fig:histcodinggain} shows the histogram of $\gamma(\Lambda_{{\bf L}^{-1}_p})$ for the same $10^5$ channel samples. There is a probability mass at around $1$, which states that unitary matrices has cleared all the bad channels with $\gamma\left(\Lambda_{\bf H}\right)<1$ to convert them to an equivalent channel matrix ${\bf L}^{-1}_p$ with $\gamma(\Lambda_{{\bf L}^{-1}_p})\geq1$. Note also that the maximum attainable coding gain (Hermite's constant~\cite{conway}) for $2$-dimensional lattices is $\frac{2}{\sqrt{3}}$. As it is clear from Figs.~\ref{fig:tanetavscodinggain} and~\ref{fig:histcodinggain}, this quantity has been achieved by few lattices in our simulations too.
\subsection{Design of Type II UPIF}
In previous Subsection, we introduced ${\bf P}_{\tiny \mbox{1,opt}}$ as a suitable precoder matrix for UPIF. Equations \eqref{ParametrizedPrecoder} and \eqref{PoptPrecoderReal} show that Type I UPIF precoder matrices ${\bf P}_{\tiny \mbox{1,opt}}$ and ${\bf P}^{(\mathbb{R})}_{\tiny \mbox{1,opt}}$ are adapted based on ${\bf L}^{-1}$ and consequently ${\bf H}$ and $\rho$. In this Subsection, however we introduce Type II UPIF, where the optimal precoder matrix ${\bf P}_{\tiny \mbox{2,opt}}$ is fixed and it does not change by varying the channel matrix ${\bf H}$.

By further expanding $\epsilon_{1}^{2}(\Lambda_{{\bf L}_p^{-1}})$ in \eqref{def:upper2}, we get:
\begin{eqnarray}
\epsilon_{1}^{2}(\Lambda_{{\bf L}_p^{-1}})&=&\|{\bf L}_p^{-1}{\bf v}\|^2\nonumber\\
&=& {\bf v}^h{\bf L}_p^{-h}{\bf L}_p^{-1}{\bf v}= {\bf v}^h\left({\bf L}_p{\bf L}_p^{h}\right)^{-1}{\bf v}\nonumber\\
&=& {\bf v}^h{\bf P}^h\left(\left({\bf I}_{2n}+\rho\cdot{\bf \Sigma}^h{\bf \Sigma}\right)^{-1}\right)^{-1}{\bf P}{\bf v}\label{equ:1}\\
&=& {\bf v}^h{\bf P}^h\left({\bf I}_{2n}+\rho\cdot{\bf \Sigma}^h{\bf \Sigma}\right){\bf P}{\bf v},\label{equ:11}
\end{eqnarray}
where \eqref{equ:1} follows from \eqref{eq:new4}. Since ${\bf I}_{2n}+\rho\cdot{\bf \Sigma}^h{\bf \Sigma}$ is a positive definite matrix, we can perform a Cholesky decomposition to obtain ${\bf I}_{2n}+\rho\cdot{\bf \Sigma}^h{\bf \Sigma} = {\bf D}{\bf D}^h$. With this, \eqref{equ:11} can be written as $\|{\bf D}{\bf P}{\bf v}\|^2$. Hence, we get
\begin{eqnarray}
\epsilon_{1}^{2}(\Lambda_{{\bf L}_p^{-1}}) &=& \|{\bf D}{\bf P}{\bf v}\|^2\nonumber\\
&=&\sum_{m=1}^{2n}\left[{\bf D}\right]_{m,m}^2\left[{\bf P}{\bf v}\right]_m^2\nonumber\\
&\geq& 2n\left(\prod_{m=1}^{2n}\left[{\bf D}\right]_{m,m}^2\prod_{m=1}^{2n}\left[{\bf P}{\bf v}\right]_m^2\right)^{\frac{1}{2n}}\label{equ:3}\\
&=& 2n\left(\prod_{m=1}^{2n}\left(1+\rho\sigma_m^2({\bf H})\right)\prod_{m=1}^{2n}\left[{\bf P}{\bf v}\right]_m^2\right)^{\frac{1}{2n}}\nonumber\\
&\geq& 2n\prod_{m=1}^{2n}\left(\rho\sigma_m^2({\bf H})\right)^{\frac{1}{2n}}\left(d_{p,\min}^2\left(\Lambda_{\bf P}\right)\right)^{\frac{1}{2n}}\label{equ:4}\\
&=& 2n\rho\left(\det({\bf H})\right)^{\frac{1}{n}}d_{p,\min}^{\frac{1}{n}}\left(\Lambda_{\bf P}\right),\label{equ:5}
\end{eqnarray}
where \eqref{equ:3} follows from Cauchy--Schwarz inequality, and \eqref{equ:4} is true because $1+\rho\sigma_m^2({\bf H})>\rho\sigma_m^2({\bf H})$, for all $1\leq m\leq 2n$ and $d_{p,\min}^2\left(\Lambda_{\bf P}\right)\leq\prod_{m=1}^{2n}\left[{\bf P}{\bf u}\right]_m^2$ for all ${\bf u}\in\mathbb{Z}^{2n}$.

Based on \eqref{equ:5} and since almost surely $\det\left({\bf H}\right)\neq0$, the optimal Type II UPIF can be obtained by solving the following optimization problem:
\begin{equation}~\label{PoptCode}
{\bf P}_{\tiny \mbox{2,opt}} = \arg\max_{{\bf P}\in\mathcal{O}_{2n}}d_{p\min}^{\frac{1}{n}}\left(\Lambda_{\bf P}\right),
\end{equation}
The solution for the maximization in \eqref{PoptCode} is provided in~\cite{OV04}--\cite{GBB97} using algebraic number theoretic lattices. A list of full-diversity algebraic rotations is available in~\cite{site}. All the lattices provided there satisfy non-vanishing minimum product distance criterion too. Hence, thanks to~\cite{site}, the optimal Type II UPIF along with their corresponding $d_{p,\min}$ are available.
\begin{figure}[t]%
\begin{center}
\includegraphics[width=9cm]{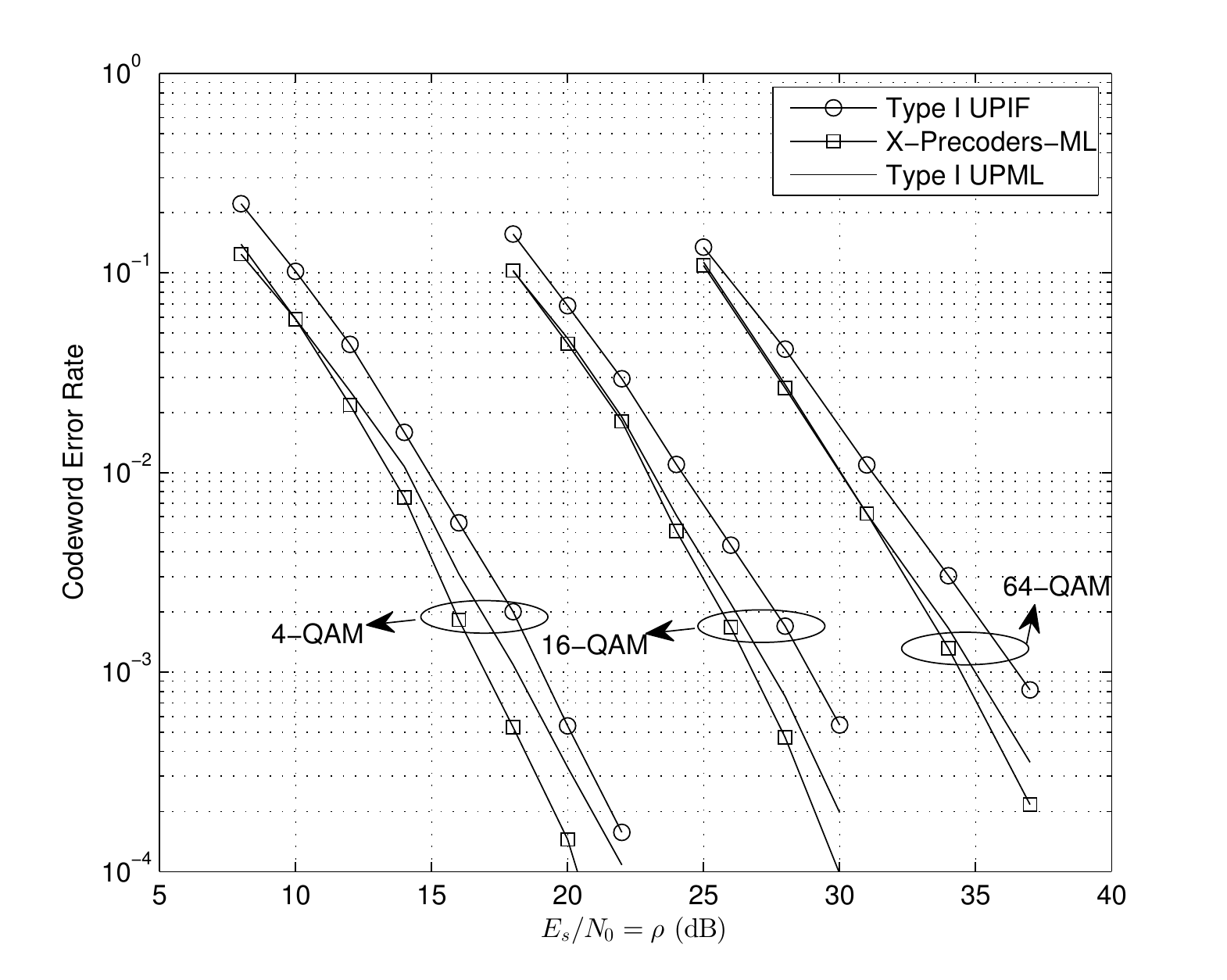}~\caption{\label{fig:22_41664_Precoders}Type I UPIF in comparison with, X-Precoders decoded with sphere decoding algorithm, and Type II UPML in a $2\times 2$ complex MIMO Channel.}
\end{center}
\end{figure}
We now proceed to verify this result by conducting simulations.
\section{Simulation Results}\label{Section:Simulations}
The UPIF scheme and MIMO precoding X-codes and Y-codes~\cite{saif11} share similar properties, which make them suitable for comparison: ({\em i}) both schemes use SVD decomposition technique to transform the channel matrix into a diagonal one, ({\em ii}) the precoder matrices in both systems must be unitary/orthogonal matrices, ({\em iii}) both the detectors at the receiver side, {\em i.e.} lattice reduction based IF linear receiver~\cite{Sakzad14-1} and a combination of two $2$-dimensional ML decoders, provide full receive diversity in $2\times 2$ MIMO.

For comparison we show the performance of UPIF precoding when the IF receiver is replaced by an ML decoder. We will denote this with the acronym UPML.
In the following subsections, we simulate and compare our proposed Type I/II UPIF decoded using lattice reduction based IF receiver, with Type I/II UPML, and MIMO X-Precoders/X-Codes.
\subsection{Type I UPIF versus X-Precoders}
We show simulations for a $2\times2$ MIMO UPIF channel over $4/16/64$-QAM constellations. The matrices ${\bf A}$ and ${\bf B}$ for IF linear receiver were found using Algorithm2 presented in~\cite{Sakzad14-1}. We have chosen our Type I UPIF by running an exhaustive search based on the design criterion given in \eqref{PoptPrecoderReal}. For reference purposes, we also presented MIMO X-precoder~\cite{saif11} scheme decoded under sphere decoding algorithm~\cite{ViB}. The step size for our brute force search for both the cases (Type I UPIF and X-precoders) is $0.001$ radians. Fig.~\ref{fig:22_41664_Precoders} shows the codeword error rate~\cite{Sakzad14-1} curves versus signal-to-noise ratio $\rho$ of the different schemes. Since the performance of UPIF decreases by growing $\rho$ parallel to ML decoded curve, full-diversity of UPIF is guaranteed.
In particular, at codeword error rate $10^{-3}$ over $4/16/64$-QAM constellations, the gaps of $2.2$dB, $2.15$dB, and $2.1$dB between Type I UPIF and X-precoder are observed. This shows a decrease in coding loss by increasing the constellation size. The $2\times 2$ X-precoders outperform the other schemes because of two main reasons: ({\em i}) these precoders are designed for the particular lattice constellation (for example, $4$-QAM, $16$-QAM, or $64$-QAM) while Type I UPIF are designed for the (infinite) lattice $\mathbb{Z}^{2n}$: ({\em ii}) since we are only able to simulate Type I UPIF over a $2\times 2$ MIMO channel, the diversity gain of Type I UPIF over X-precoders cannot be captured. This will be more visible in the next subsection, where Type II UPIF outperform X-codes in MIMO channels with larger number of antennas employed at larger constellation sizes.
\subsection{Type II UPIF versus X-Codes}
\begin{figure}[t]%
\begin{center}
\includegraphics[width=9cm]{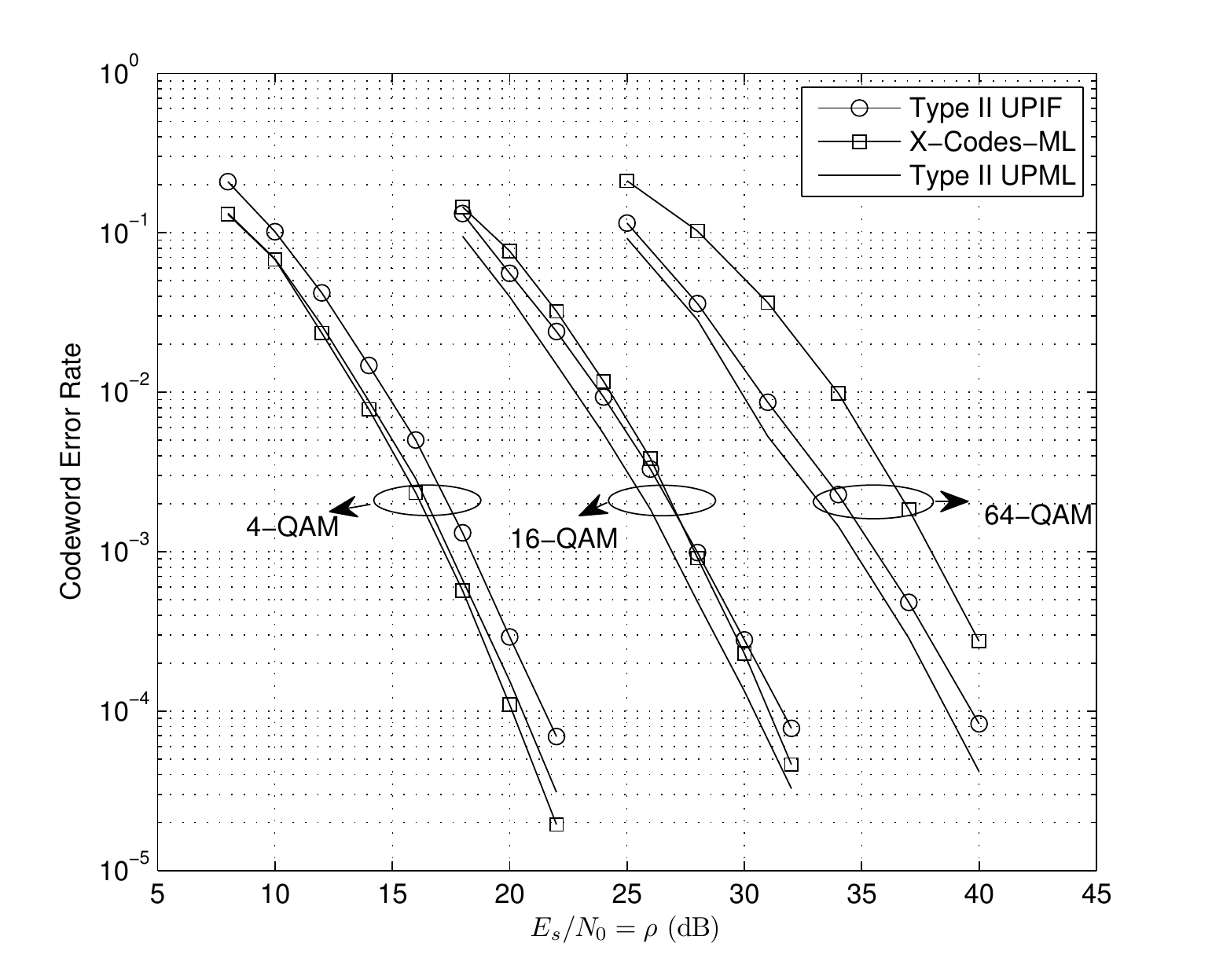}~\caption{\label{fig:22_41664_Codes}Type II UPIF in comparison with X-Codes and Type II UPML in a $2\times 2$ complex MIMO Channel.}
\end{center}
\end{figure}

\begin{figure}[t]%
\begin{center}
\includegraphics[width=9cm]{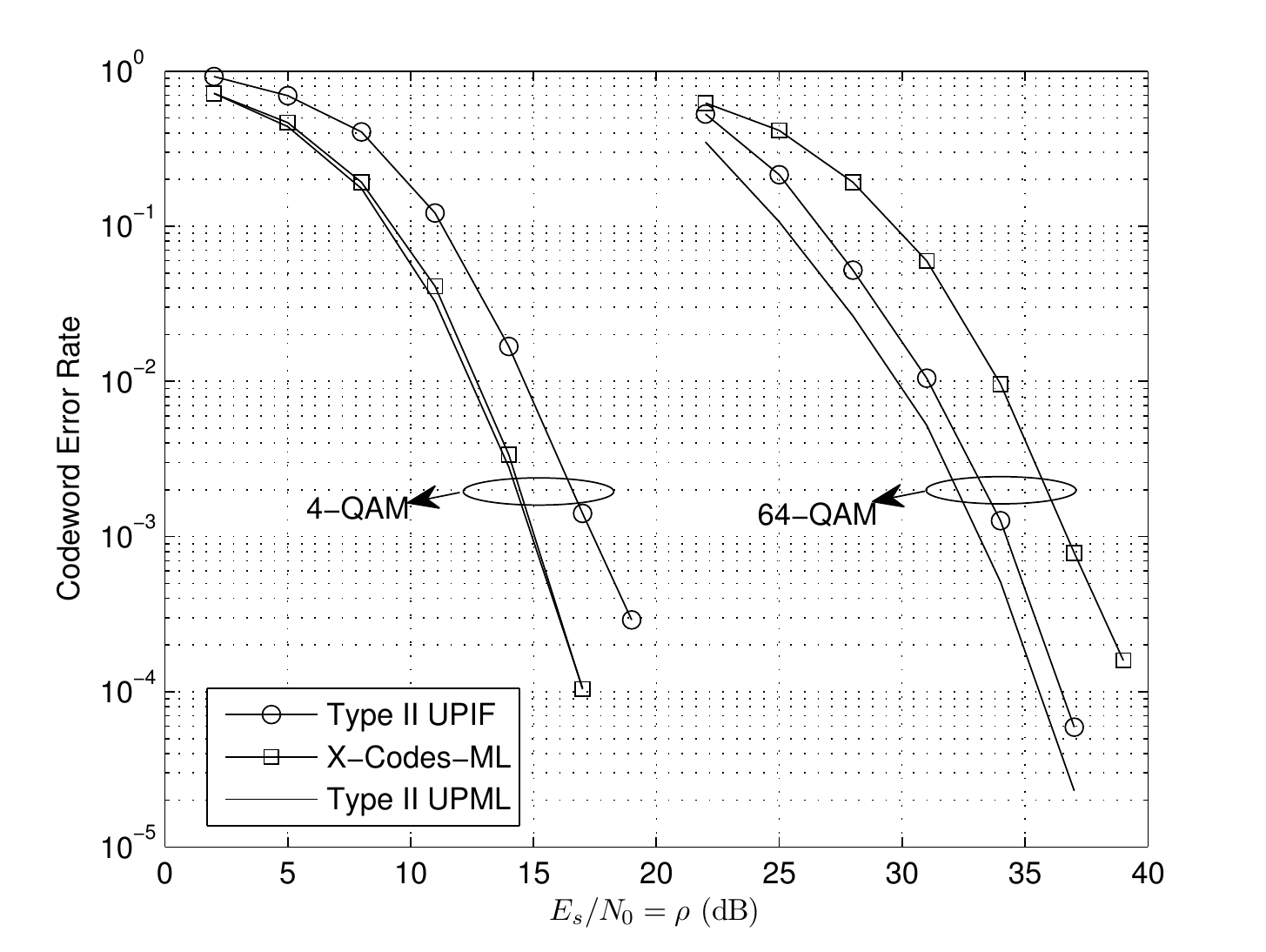}~\caption{\label{fig:44_464_Codes}Type II UPIF in comparison with X-Codes and Type II UPML in a $4\times 4$ complex MIMO Channel.}
\end{center}
\end{figure}

We have conducted simulations for $2\times2$ and $4\times 4$ MIMO UPIF channel over $4/16/64$-QAM constellations. For $2\times 2$ MIMO channels, the matrices ${\bf A}$ and ${\bf B}$ for IF linear receiver were found using Algorithm2 presented in~\cite{Sakzad14-1} while for larger MIMOs we have employed Algorithm1. We have chosen our Type II UPIFs based on the design criterion given in \eqref{PoptCode}. The $2\times 2$ and $4\times 4$ full-diversity algebraic rotation~\cite{site} matrices which has been used for simulations have the following minimum product distances ($d^{\frac{1}{n}}_{p,\min}$)
$0.668740$, $0.438993$, and $0.289520$. For reference purposes, we also presented MIMO X-codes~\cite{saif11} decoded under sphere decoding algorithm~\cite{ViB}.
For $4$-QAM constellation the optimal angle for a X-Code is $26.6$ degrees while for $16$-QAM it is equal to $15$ degrees~\cite{saif11}. We also found the optimal angle for $64$-QAM to be equal to $8$ degrees.
In Figs.~\ref{fig:22_41664_Codes} and~\ref{fig:44_464_Codes}, we plot the codeword error performance of Type II UPIF versus MIMO X-codes and Type II unitary precoders decoded using a maximum likelihood decoder (Type II UPML) such as sphere decoding algorithm versus signal-to-noise ratio $E_s/N_0=\rho$.
As it is clear from Figs.~\ref{fig:22_41664_Codes} and~\ref{fig:44_464_Codes}, the Type II UPIF outperform MIMO X-codes for larger constellation sizes in higher MIMO dimensions. The main reason is that the Type II UPIF precoder matrices are designed to achieve full-diversity and the diversity gain shows itself when the number of antennas increases. In particular, the diversity order~\cite{saif11} of the optimal MIMO X-codes (in terms of diversity not the error performance) is known to be $d_{n_t,n_r,n_s}=\left(n_r-\frac{n_s}{2}+1\right)\left(n_t-\frac{n_s}{2}+1\right)$, where $n_r$ and $n_t$ denote the number of receive and transmit antennas, respectively, and $n_s$ denotes the number of symbols. In our setting, for $2\times 2$ and $4\times 4$ MIMO X-codes, we have $d_{2,2,2}=4$ and $d_{4,4,4}=9$ in comparison with full-diversity orders $4$ and $16$ for Type II UPIF, respectively.
\subsection{Type I UPIF versus Type II UPIF}
\begin{figure}[t]%
\begin{center}
\includegraphics[width=9cm]{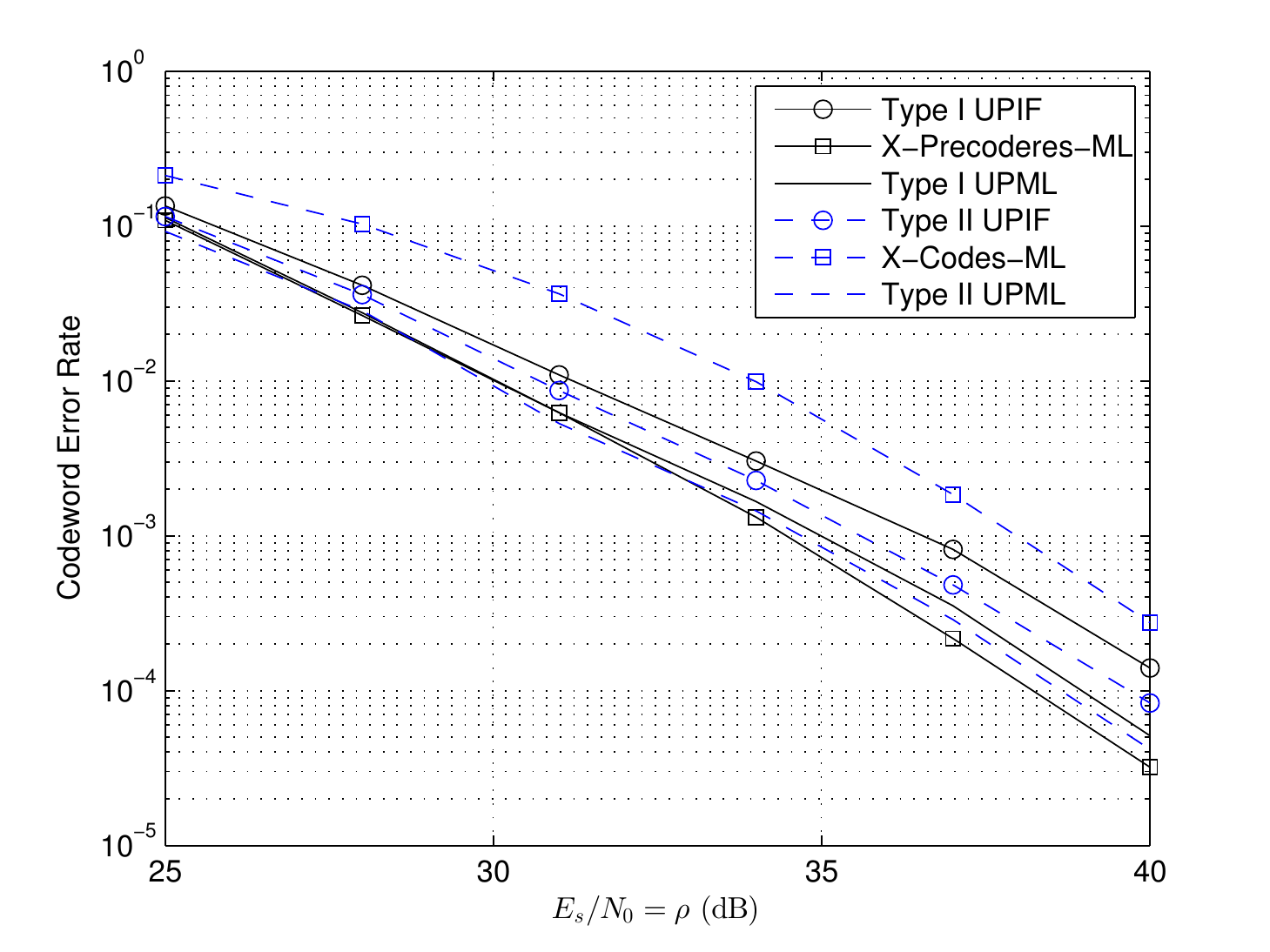}~\caption{\label{fig:22_64_PrecodersCodes}Type I versus Type II UPIF and UPML schemes in a $2\times 2$ complex MIMO Channel.}
\end{center}
\end{figure}

In this subsection, we compare Type I and Type II UPIF together. Fig.~\ref{fig:22_64_PrecodersCodes} shows the codeword error performance of Type I and Type II UPIF in comparison to their corresponding UPML. It is obvious that the Type II UPIF and UPML outperform Type I UPIF and UPML precoders by no more than $0.5$dB and $1$dB, respectively.

The sub-optimality of unitary precoder matrices obtained using quantized exhaustive search algorithm is the main reason for having Type II UPIF performing better than Type I UPIF. Furthermore, we note that the design criteria given in \eqref{PoptPrecoderReal} and \eqref{PoptCode} are different and we cannot say which one can perform better than the other by just looking at these equations. Hence, considering the simplicity of Type II UPIF in comparison to Type I and their better error performance, we suggest using of Type II UPIF for practical purposes.
\section{Summary and Directions for Future Work}~\label{Section:Conclusion}
A unitary precoding scheme (UPIF) has been introduced for transmission over a flat-fading MIMO channel with  both CSIT and CSIR, and an IF linear receiver.
The diversity gains of the proposed approach have been analyzed both theoretically and numerically. Two different design criteria for two types of unitary precoders, Type I UPIF and Type II UPIF, were given. For both cases, an SVD is performed first to transform the channel to a diagonal one. Type I UPIF includes precoder matrices adapting based on the channel matrix singular values, while Type II UPIF consists of precoding matrices, which remain fixed by changing the channel matrix.

Designing full-diversity unitary precoders or more generally space-time block codes for IF receivers without CSIT would be of interest. Another direction is to let the transmitter have access to limited feedback over a delay-free link from the IF receiver~\cite{sakzad14,sakzad14-ISIT}. In this case, designing a suitable codebook of unitary precoding matrices attaining higher rates and obtaining higher coding gains in error performance seems to be a promising research topic.

\appendices
\section{Proof of Theorem~\ref{th:ub}}~\label{app:upperboundproof}
For the proof of this we follow the lines of the proof of Theorem 7 in \cite{Feng}. Since the minimum Euclidean distance of $\mathbb{Z}$ is unity, an error is declared if ${\bf e}_m \geq \frac{\sqrt{\rho}}{2}$. Therefore, $P_{e}\left(m, {\bf \Sigma}{\bf P}, \mathbb{Z}^{2n}\right)$ equals to
\begin{eqnarray}
\!\!\!\!\!\!\!\!\!\!\!\!\!\!\!\!&=&\!\!\!\!\!\! \mbox{Pr}\left(\left|{\bf e}_m\right| \geq \frac{\sqrt{\rho}}{2}\right) = 2\mbox{Pr}\left({\bf e}_m \geq \frac{\sqrt{\rho}}{2}\right)\label{P_e_0}\\
\!\!\!\!\!\!\!\!\!\!\!\!\!\!\!\!&\leq&\!\!\!\!\!\! 2\min_{t>0}\frac{\mathbb{E}(\exp(t{\bf e}_m))}{\exp\left(\frac{\sqrt{\rho} t}{2}\right)}\label{P_e_1}\\
\!\!\!\!\!\!\!\!\!\!\!\!\!\!\!\!&=&\!\!\!\!\!\! 2\min_{t>0}\frac{\mathbb{E}(\exp\left(t\sqrt{\rho}\cdot\!\langle{\bf b}_m{\bf \Sigma}{\bf P}\!-\!{\bf a}_m,{\bf x}_m\rangle\!+\!t\cdot\!\langle{\bf b}_m,{\bf z}'_m\rangle\right))}{\exp\left(\frac{\sqrt{\rho} t}{2}\right)},\label{eq:inter}
\end{eqnarray}
where \eqref{P_e_0} follows from the symmetry of effective noise around zero, \eqref{P_e_1} follows from Chernoff's bound. Since ${\bf x}_m$ and ${\bf z}_m'$ are independent, the equation \eqref{eq:inter} can be written as
\begin{equation}~\label{P_e_2}
\min_{t>0}\frac{\mathbb{E}(\exp(t\sqrt{\rho}\cdot\!\langle{\bf b}_m{\bf \Sigma}{\bf P}\!-\!{\bf a}_m,{\bf x}_m\rangle))\mathbb{E}(\exp\left(t\cdot\!\langle{\bf b}_m,{\bf z}'_m\rangle\right))}{\frac{1}{2}\exp\left(\frac{\sqrt{\rho} t}{2}\right)}.
\end{equation}
We continue to use moment generating functions of uniform and Gaussian distributions to further upper bound the two exponential terms in \eqref{P_e_2}. Since the components of ${\bf z}_m'$ are distributed based on a Gaussian distribution with zero mean and unit variance, we have
\begin{equation}~\label{eq:guass}
\mathbb{E}\left(\exp\left(t\cdot\langle{\bf b}_m,{\bf z}'_m\rangle\right)\right)\leq\exp\left(\frac{t^2\|{\bf b}_m\|^2}{2}\right),
\end{equation}
for the second term in \eqref{P_e_2}. We now investigate its first term. Let ${\bf q}_m \triangleq t\sqrt{\rho}\cdot({\bf b}_m{\bf \Sigma}{\bf P}-{\bf a}_m)$. The entries of ${\bf x}_m\in\mathbb{Z}^{2n}$ are uniformly distributed over the hypercube $\left[-t\sqrt{\rho}\|{\bf q}_m\|,t\sqrt{\rho}\|{\bf q}_m\|\right]^{2n}$, therefore $\mathbb{E}\left(\exp(t\sqrt{\rho}\cdot\langle{\bf q}_m,{\bf x}_m\rangle)\right)$ equals to
\begin{eqnarray}
&=&\mathbb{E}\left(\exp\left(t\sqrt{\rho}\cdot\sum_{j=1}^{2n}[{\bf q}_{m}]_j[{\bf x}_{m}]_j\right)\right)\nonumber\\
&=&\prod_{j=1}^{2n}\mathbb{E}\left(\exp\left(t\sqrt{\rho}\cdot[{\bf q}_{m}]_j[{\bf x}_{m}]_j\right)\right)~\label{eq:uniform}\\
&\leq&\prod_{j=1}^{2n}\frac{\sinh\left(t\sqrt{\rho}|[{\bf q}_{m}]_j[{\bf x}_{m}]_j|\right)}{t\sqrt{\rho}|[{\bf q}_{m}]_j[{\bf x}_{m}]_j|}~\label{eq:uniform1}\\
&\leq&\prod_{j=1}^{2n} \exp\left(\frac{t^2\rho |[{\bf q}_{m}]_j|^2}{6}\right)\label{eq:uniform2}\\
&\leq& \exp\left(\frac{t^2\rho \|{\bf q}_m\|^2}{2}\right)\label{eq:uniform3},
\end{eqnarray}
where \eqref{eq:uniform} follows from the independence of components of ${\bf x}_m$, \eqref{eq:uniform1} follows from the moment generating function of a uniform distribution, and \eqref{eq:uniform2} follows from the inequality
$$\frac{\sinh(y)}{y}\leq \exp\left(\frac{y^2}{6}\right),$$
for every $y\in\mathbb{R}$. Combining \eqref{eq:uniform3} and \eqref{eq:guass} in \eqref{P_e_2}, we get $P_{e}(m, {\bf \Sigma}{\bf P}, \mathbb{Z})$ less than or equal to
\begin{eqnarray}
&\leq&2\min_{t>0}\frac{\exp\left(\frac{t^2\rho\|{\bf b}_m{\bf \Sigma}{\bf P}-{\bf a}_m\|^2}{2}\right)\exp\left(\frac{t^2\|{\bf b}_m\|^2}{2}\right)}{\exp\left(\frac{\sqrt{\rho} t}{2}\right)}\nonumber\\
&=&2\min_{t>0}\frac{\exp\left(\frac{t^2G({\bf a}_m,{\bf b}_m)}{2}\right)}{\exp\left(\frac{\sqrt{\rho} t}{2}\right)}\label{P_e_3}\\
&=&2\exp\left(\frac{-\rho}{4G({\bf a}_m,{\bf b}_m)}\right),\label{P_e_4}
\end{eqnarray}
where \eqref{P_e_4} obtained by optimizing $t$, that is $t=\frac{\sqrt{\rho}}{2G({\bf a}_m,{\bf b}_m)}$.

Now since $G({\bf a}_m,{\bf b}_m)$ needs to be minimized, ${\bf a}_{m}$ and ${\bf b}_{m}$ should be chosen appropriately as in \cite{zhan12} and \cite{Sakzad14-1}, thus we get
$$\frac{G({\bf a}_m,{\bf b}_m)}{\rho}=\epsilon_{m}^{2}(\Lambda_{{\bf L}_p^h}),$$
where $\epsilon_{m}^{2}(\Lambda_{{\bf L}_p^h})$ denotes the $m$-th successive minimum of the lattice
$$\Lambda_{{\bf L}_p^h} = \left\lbrace {\bf L}^h{\bf P}{\bf d} ~|~ \forall {\bf d} \in \mathbb{Z}^{2n} \right\rbrace.$$
Here ${\bf L}_p^{-1}$ is a generator of the dual lattice $\Lambda_{{\bf L}_p^h}^\ast$. Thus we have the relation (see Lemma $4$ in \cite{zhan12})
\begin{equation}
\label{bound2}
\epsilon_{m}^{2}(\Lambda_{{\bf L}_p^h}) \leq \frac{(2n)^{3} + (3n)^{2}}{\epsilon_{2n-m+1}^{2}(\Lambda_{{\bf L}_p^h}^\ast)}=\frac{(2n)^{3} + (3n)^{2}}{\epsilon_{2n-m+1}^{2}(\Lambda_{{\bf L}_p^{-1}})},
\end{equation}
where $\epsilon_{2n-m+1}^{2}(\Lambda_{{\bf L}_p^h}^\ast)$ is the $(2n-m+1)$-th successive minima of the lattice $\Lambda_{{\bf L}_p^h}^\ast$. Therefore, we have
\begin{equation}\label{bound1}
\frac{\rho}{G({\bf a}_m,{\bf b}_m)}\geq \frac{\epsilon_{2n-m+1}^{2}(\Lambda_{{\bf L}_p^{-1}})}{c_0},
\end{equation}
where $c_0 = (2n)^{3} + (3n)^{2}$.
Using inequality of \eqref{bound1} in \eqref{P_e_4}, the probability of error for decoding the $m$-th layer is upper bounded as
\begin{equation}
P_{e}\left(m, {\bf \Sigma}{\bf P}, \mathbb{Z}^{2n}\right) \leq \exp\left(-c\epsilon_{2n-m+1}^{2}(\Lambda_{{\bf L}_p^{-1}})\right),
\end{equation}
where $c = \frac{1}{4c_0}$. This completes the proof.
\section{Proof of Theorem~\ref{th:diversity}}~\label{app:diversityproof}
Toward the proof of this theorem, we make use of the following two lemmas. The first lemma gives the expression for the first-order expansion of the marginal probability density function (PDF) of eigenvalues of ${\bf H}$.
\begin{lemma}~\label{lemma:eigen}
Let the entries of the matrix ${\bf H}$ be i.i.d. complex Gaussian random variables with zero mean and
unit variance. The first-order expansion of the marginal PDF of
the $k$-th largest eigenvalue $\lambda_k$ of the complex central Wishart matrix ${\bf H}{\bf H}^h$
is given by
$$F_{\lambda_k}(\lambda_k)=q_k\lambda_k^{d_k}+o\left(\lambda_k^{d_k}\right),$$
as $\lambda_k\rightarrow0$, with $d_k=(2n-k+1)^2-1$ and $q_k$ being positive constants.
\end{lemma}
\begin{lemma}\label{lemma:Pe}
For a scalar channel modeled by $y=\sqrt{\mbox{\small SNR}}\beta x + n$ , where $n\sim\mathcal{N}(0,1)$, and $\mathbb{E}[|x|^2]=1$ and $\alpha = \beta^2$
is a nonnegative random variable whose probability density
function (PDF) $F_\alpha(\alpha)$ is such that as
\begin{equation}~\label{eq:PDFsing}
F_\alpha(\alpha) = q\alpha^t + o(\alpha^t),
\end{equation}
the average word error probability (WEP) $P_e$, which is given by
$$P_e=\mathbb{E}\left(P_{e}(\alpha)\right)=\int_0^\infty \exp\left(-k\alpha\mbox{\small SNR}\right)F_\alpha(\alpha)d\alpha,$$ is
such that as $\mbox{\small SNR}\rightarrow\infty$
$$P_e=\frac{2^tq\Gamma(t+\frac{3}{2})}{\sqrt{\pi}(t+1)}(k\mbox{\small SNR})^{-(t+1)}+o\left(\mbox{\small SNR}^{-(t+1)}\right).$$
\end{lemma}
Now we proceed to give the proof of Theorem~\ref{th:diversity}. Let ${\bf v}$ be the Gaussian vector for which
\begin{eqnarray}
\epsilon_{1}^{2}(\Lambda_{{\bf L}_p^{-1}})&=&\|{\bf L}_p^{-1}{\bf v}\|^2\nonumber\\
&=& {\bf v}^h{\bf L}_p^{-h}{\bf L}_p^{-1}{\bf v}= {\bf v}^h\left({\bf L}_p{\bf L}_p^{h}\right)^{-1}{\bf v}\nonumber\\
&=& {\bf v}^h{\bf P}^h\left(\left({\bf I}_{2n}+\rho\cdot{\bf \Sigma}^h{\bf \Sigma}\right)^{-1}\right)^{-1}{\bf P}{\bf v}\label{eq:1}\\
&=& {\bf v}^h{\bf P}^h\left({\bf I}_{2n}+\rho\cdot{\bf \Sigma}^h{\bf \Sigma}\right){\bf P}{\bf v}\nonumber\\
&\triangleq& \|{\bf D}{\bf P}{\bf v}\|^2 \geq \left[{\bf D}\right]_{1,1}^2\left[{\bf P}{\bf v}\right]_1^2\label{eq:2}\\
&=& (1+\rho\sigma_1^2({\bf H}))\left[{\bf P}{\bf v}\right]_1^2\geq\rho\sigma_1^2({\bf H})\left[{\bf P}{\bf v}\right]_1^2\label{eq:3}\\
&\geq&\sqrt{\rho\lambda_1({\bf H})}\left[{\bf P}{\bf v}\right]_1^2\label{eq:4}
\end{eqnarray}
where \eqref{eq:1} follows from \eqref{eq:new4}, \eqref{eq:2} follows from Cholesky decomposition of ${\bf I}_{2n}+\rho\cdot{\bf \Sigma}^h{\bf \Sigma}$ and the fact that ${\bf D}$ is a diagonal matrix, \eqref{eq:3} is true because $1+\rho\sigma_1^2({\bf H})>\rho\sigma_1^2({\bf H})$, and \eqref{eq:4} follows from the fact that $\sigma_1^2({\bf H})=\lambda_1({\bf H})$.

Since the quantity $[{\bf P}{\bf v}]_1$ is not zero. Using Lemmas~\ref{lemma:eigen},~\ref{lemma:Pe} and \eqref{eq:4}, we have
$$P_e \triangleq \mathbb{E}_{\bf H}\left(P_{e}({\bf \Sigma}{\bf P}, \mathbb{Z}^{2n})\right) \leq t_0\rho^{-(2n)^2}+o\left(\rho^{-(2n)^2}\right),$$
where $t_0$ is a function of $n$, $q_1$ defined in \eqref{eq:PDFsing}, and $[{\bf P}{\bf v}]_1\neq0$.

\end{document}